\documentclass[
superscriptaddress,
twocolumn,
amsmath,amssymb,
aps,
prc
]{revtex4-1}

\usepackage{graphicx}
\usepackage{dcolumn}
\usepackage{bm}
\usepackage{soul,xcolor}
\usepackage{mathrsfs}
\usepackage[fontset=none]{ctex}

\usepackage{ulem}
\setstcolor{blue}
\usepackage{longtable}
\usepackage{geometry}

\usepackage[pdfencoding=auto,psdextra]{hyperref}

\usepackage{setspace}

\usepackage{geometry}
\usepackage{mathtools}

\begin{document}
	\begin{CJK*}{UTF8}{}
		
		\title{Shell structure and spontaneous decay of superheavy $Z=124$ isotopes with the deformed relativistic Hartree-Bogoliubov theory in continuum}

	\author{Yumeng Wang (\CJKfamily{gbsn}王宇萌)}
	\affiliation{School of Optoelectronic Information and Physical Science, Jiangnan University, Wuxi 214122, China}
	
	\affiliation{
			Jiangnan Institute of Advanced Study in Wuxi Liangxi,
			Wuxi 214021, China}
	
\affiliation{
		Wuxi Key Laboratory of Optoelectronic Intelligent Perception, Wuxi 214122, China}
		
	\affiliation{
		Jiangsu Provincial Research Center of Light Industrial Optoelectronic Engineering and Technology, Wuxi 214122, China}

	\author{Lang Liu (\CJKfamily{gbsn}刘朗)}
	\email{liulang@jiangnan.edu.cn}
	\affiliation{School of Optoelectronic Information and Physical Science, Jiangnan University, Wuxi 214122, China}
	
	\affiliation{
	Jiangnan Institute of Advanced Study in Wuxi Liangxi,
		Wuxi 214021, China}

\affiliation{
		Wuxi Key Laboratory of Optoelectronic Intelligent Perception, Wuxi 214122, China}

\affiliation{
		Jiangsu Provincial Research Center of Light Industrial Optoelectronic Engineering and Technology, Wuxi 214122, China}

	\author{Cong Pan (\CJKfamily{gbsn}潘琮)}
	\email{cpan@ahnu.edu.cn}
	\affiliation{Department of Physics, Anhui Normal University, Wuhu 241000, China}

		\renewcommand{\abstractname}{Abstract}
		
		\begin{abstract}

	Shell structure and spontaneous decay of $Z = 124$ isotopes are systematically investigated using the deformed relativistic Hartree-Bogoliubov theory in continuum (DRHBc) with the PC-PK1 density functional. By accounting for deformation and continuum effects, the neutron drip line is predicted to be located at $N = 320$, and eight potential multi-neutron emitters beyond the neutron drip line are identified.
	The results, compared with the spherically constrained case, highlight the critical role of deformation effects in determining the ground-state properties of superheavy nuclei. 
	In addition, potential neutron shell closures emerge at $N = 258$ and $350$, alongside subshell closures at $N = 232$ and $320$. The results disfavor $Z = 124$ as a possible proton magic number and suggest $Z = 120$ and $138$ as more favorable candidates for proton shell closures in the superheavy region. Finally, the competition between $\alpha$ decay, $\beta$ decay, and spontaneous fission is analyzed by calculating half-lives within various semiempirical formulas. 
	The results indicate that $\alpha$ decay is the predominant mode on the proton-rich side, whereas spontaneous fission and $\beta^-$ decay gradually become dominant with increasing neutron number.
			
			\end{abstract}

		\maketitle
	\end{CJK*}

	\section{Introduction}
	\label{sec:1}
	The synthesis and stability of superheavy elements are frontier topics in nuclear physics
	~\cite{Cwiok1996NPA611,Myers1966NP81,Erler2012N486,Thoennessen2013RoPiP76,Moeller2016EWC131}. To date, superheavy elements with proton numbers up to $Z=118$ have been successfully synthesized experimentally through cold and hot fusion reactions~\cite{Oganessian1999N400,Oganessian2000PRC63,Morita2004JPSJ73,Oganessian2006PRC74,Morita2007JPSJ76,Oganessian2010PRL104,Oganessian2013PRC87}, thereby completing the seventh period of the periodic table. The exploration of elements beyond $Z=118$, such as $Z=119$ and $120$, is still ongoing~\cite{Oganessian2009PRC79,Hofmann2016EPJA52,Khuyagbaatar2020PRC102,Sakai2022EPJA58,Tanaka2022JPSJ91,Gan2022EPJA58,Zhang2025NST36}. Although next-generation radioactive ion beam facilities are evolving rapidly, the majority of neutron-rich nuclei far from the valley of stability are expected to remain beyond the reach of experiments in the foreseeable future. Therefore, reliable theoretical models are essential for investigating the properties of unknown nuclei and providing guidance for experiments.
	
    Various theoretical approaches have been developed to describe nuclear masses and ground-state properties. Hybrid ``macroscopic-microscopic'' models include the finite-range droplet model (FRDM) ~\cite{Moeller1995ADNDT59, Moeller2016ADNDT109},
    the extended Thomas-Fermi plus Strutinsky integral method~\cite{Aboussir1995ADNDT61,Pearson1996PLB387}, and the Weizsäcker-Skyrme (WS) model~\cite{Wang2010PRC82,Wang2010PRC82}. Within the nonrelativistic framework, density functional theories based on Skyrme~\cite{Samyn2002NPA700,Stoitsov2003PRC68,Goriely2009Prl102,Erler2012N486,Goriely2013PRC88} and Gogny~\cite{Hilaire2007EPJA33,Goriely2009PRL102a,Delaroche2010PRC81} interactions have been widely applied. In the relativistic framework, extensive studies based on the covariant density functional theory (CDFT) have been performed~\cite{Lalazissis1999ADNDT71,Geng2005Potp113,Zhang2014FP9,Agbemava2014PRC89,Lu2015PRC91,Xia2018ADNDT121122} and have attracted wide attention.

    CDFT has been proved to be a powerful theory in nuclear physics due to its successful description of many nuclear phenomena~\cite{Meng2016,Ring1996PPNP37,Vretenar2005PR409,Meng2006PPNP57,Niksic2011PPNP66,Meng2013FP8,Meng2015JPGPP42,Zhou2016PS91,Shen2019PiPaNP109}. Based on CDFT, the relativistic continuum Hartree-Bogoliubov (RCHB) theory was developed with a microscopic and self-consistent treatment of pairing correlations and continuum effects~\cite{Meng1996Prl77,Meng1998NPA635}. The RCHB theory has successfully described many properties of both stable and exotic nuclei, including the halo structure in $^{11}$Li~\cite{Meng1996Prl77}, predictions of giant halo phenomena~\cite {Meng1998Prl80,Meng2002PRC65,Zhang2003SCSG46}, halos in hypernuclei~\cite{Lue2003EPJA17}, explanations of pseudospin symmetry in exotic nuclei~\cite{Meng1998PRC58,Meng1999PRC59}, and predictions of new magic numbers in superheavy nuclei~\cite{Zhang2005NPA753}. Based on the RCHB theory, the first nuclear mass table including continuum effects was constructed for nuclei with $8 \le Z \le 120$~\cite{Xia2018ADNDT121122}. 
    Considering that most nuclei are deformed, while the RCHB theory assumes spherical symmetry, its accuracy in describing nuclear properties such as masses is limited for nuclei away from magic numbers. 
    
    To properly account for deformation effects, the deformed relativistic Hartree-Bogoliubov theory in continuum (DRHBc) was subsequently developed~\cite{Zhou2010PRC82,Li2012PRC85}. This theory solves the axially relativistic
    Hartree-Bogoliubov (RHB) equations on the Dirac Woods-Saxon basis~\cite{Zhou2003PRC68,Zhang2022PRC106}, taking into account deformation effects, pairing correlations, and continuum effects, thereby providing proper descriptions of the ground-state properties of deformed nuclei. 
    The DRHBc theory has been successfully applied to many studies of exotic nuclei, such as descriptions of halo nuclei~\cite{Zhou2010PRC82,Li2012PRC85,Sun2018PLB785,Zhang2019PRC100,Sun2020NPA1003,Yang2021PRL126,Sun2021PRC103,Sun2021SB66,Zhong2022SCPMA65,Zhang2023PLB844,Zhang2023PRC107,Wang2024EPJA60,Pan2024PLB855,An2024PLB849,Zhang2024PRC110a,Papakonstantinou2025PRC112,Zhang2025PLB871,Zhang2026P9,Pan2026PLB877,Lu2026PRC114}, dripline locations~\cite{In2021IJMPE30,Zhang2021PRC104,Pan2021PRC104,He2024PRC110}, evolution of shell structures~\cite{Zheng2024CPC48,Zhang2024PRC110,Huang2025PRC111,Zhang2025PRC112} and deformation~\cite{Mun2023PLB847,Guo2023PRC108}, shape coexistence~\cite{In2020JKPS77,Choi2022PRC105,Kim2022PRC105,Pan2025PRC112}, and fission barriers~\cite{Zhang2024CPC48}.
     The nuclear mass table in the DRHBc theory for even-even nuclei~\cite{Zhang2020PRC102} and even-$Z$ nuclei~\cite{Pan2022PRC106} with $8 \le Z \le 120$ has been established~\cite{Zhang2022ADNDT144,Guo2024ADNDT158,Zhang2025AB35}.
    The nuclear mass table within the DRHBc theory for odd-$Z$ nuclei with $8 \le Z \le 120$ is currently under construction.
    In addition, superheavy nuclei with $121 \le Z \le 136$ have recently been investigated within the DRHBc framework. The specific criterion for determining the ground states of superheavy nuclei is presented in Refs.~\cite{Wang2024P7,Zhuang2026CPC}.

    Shell effects are crucial for the stability of superheavy nuclei, which lead to the formation of the so-called island of stability~\cite{Myers1966NP81,Sobiczewski1966PL22,Meldner1966AfF36}. 
    The identification of proton magic numbers in the superheavy region has been a primary question in nuclear physics, with previous studies suggesting possible candidates at $Z=114,120,124,126$, and $138$~\cite{Moeller1994JPGNPP20,Cwiok1996NPA611,Denisov2005PAN68,Zhang2005NPA753,Ismail2015JPGPP43,Asous2025IJMPE34}. 
    Motivated by this, systematic studies of the ground-state properties along the $Z = 124$ isotopic chain can provide valuable insights into whether $Z = 124$ could be regarded as a proton magic number in the superheavy region.
    Furthermore, as the two dominant decay modes of superheavy nuclei, $\alpha$ decay and spontaneous fission (SF) can be taken as the limiting factor that determines the stability of the heaviest nuclei~\cite{Qian2014PRC90}.
    Therefore, theoretical calculations of $\alpha$ decay energies and spontaneous fission barriers and estimations of their half-lives can provide important guidance for experimental investigations~\cite{Denisov2024PRC110, Denisov2025ADNDT161}.

    In the present work, we perform a systematic study of the ground-state properties of the $Z=124$ isotopes within DRHBc.
    Notably, the element with atomic number $Z = 124$ is currently designated as the temporary systematic IUPAC name unbiquadium and the symbol Ubq~\cite{Chatt1979PAC51}.
     We first determine the dripline location of this isotopic chain and analyze its bulk properties. Next, potential neutron (sub)shell closures and proton shell closures in the superheavy region are examined. Finally, by calculating the half-lives associated with different decay modes, the dominant decay modes of this isotopic chain are predicted.
    This paper is organized as follows. Sec.~\ref{sec:2} briefly introduces the theoretical framework. Numerical details are presented in Sec.~\ref{sec:3}. The results and discussion are given in Sec.~\ref{sec:4}. The last section is the summary.

	\section{Theoretical framework}
	\label{sec:2}
	
	The detailed formalism of the DRHBc theory can be found in Refs.~\cite{Zhou2010PRC82,Li2012PRC85,Zhang2020PRC102}. Here, we briefly present its formalism.
	
	In the DRHBc theory, the RHB equation~\cite{Kucharek1991ZfPAHaN339}, which can self-consistently handle the mean field and pairing correlation, is expressed as
	\begin{equation}
		\begin{pmatrix}
			\hat{h}_D - \lambda_\tau & \hat{\Delta} \\
			-\hat{\Delta}^* & -\hat{h}_D^* + \lambda_\tau
		\end{pmatrix}
		\begin{pmatrix}
			U_k \\
			V_k
		\end{pmatrix}
		= E_k
		\begin{pmatrix}
			U_k \\
			V_k
		\end{pmatrix},
		\label{eq:1}
	\end{equation}  
	where $\lambda_{\tau}$ is the Fermi energy (${\tau}$ = $n$ or $p$ for neutrons or protons), $E_k$ is the quasiparticle energy, $U_k$ and $V_k$ are the quasiparticle wave functions, and $\hat{h}_D$ is the Dirac Hamiltonian.
	
	In the coordinate space, 
	\begin{equation}
		h_D(\boldsymbol{r}) = \boldsymbol{\alpha} \cdot \boldsymbol{p} + V(\boldsymbol{r}) + \beta[M + S(\boldsymbol{r})],
		\label{eq:2}
	\end{equation} 
	where $M$ is the nucleon mass, and $S(\boldsymbol{r})$ and $V(\boldsymbol{r})$ are the scalar and vector potentials, respectively.
	
	The pairing potential $\Delta$ reads 
	\begin{equation}
		\Delta(\boldsymbol{r}_1, \boldsymbol{r}_2) = V^{\text{pp}}(\boldsymbol{r}_1, \boldsymbol{r}_2) \kappa(\boldsymbol{r}_1, \boldsymbol{r}_2),
		\label{eq:3}
	\end{equation}
	where $\kappa$ is the pairing tensor~\cite{Ring1980} and the density-dependent zero-range pairing force $V^{\text{pp}}$ is
	\begin{equation}
		V^{pp}(\boldsymbol{r}_1, \boldsymbol{r}_2) = \frac{1}{2} V_0 (1 - P^\sigma) \delta(\boldsymbol{r}_1 - \boldsymbol{r}_2) \left(1 - \frac{\rho(\boldsymbol{r}_1)}{\rho_{\text{sat}}}\right).
		\label{eq:4}
	\end{equation}

	For axially deformed nuclei, the potentials and densities are
	expanded in terms of the Legendre polynomials,
	\begin{equation}
		f(\bm{r}) = \sum_{\lambda} f_\lambda(r) P_\lambda(\cos\theta), \ \lambda = 0, 2, 4, \ldots, \lambda_{\text{max}} ,
		\label{eq:5}
	\end{equation}
	where $\lambda$ is restricted to be even numbers due to spatial reflection symmetry and 
	\begin{equation}
		f_\lambda(r) = \frac{2\lambda + 1}{4\pi} \int d\Omega f(\boldsymbol{r}) P_\lambda(\Omega).
		\label{eq:6}
	\end{equation}
	
	 To properly consider the continuum effects, the deformed RHB equations 
	are solved in a spherical Dirac Woods-Saxon basis, 
	whose radial wave function has a proper asymptotic
	behavior at large $r$~\cite{Zhou2003PRC68}.

	\section{Numerical details}
	\label{sec:3}
	
	Numerical details for the DRHBc calculations of $\prescript{ }{124}{\text{Ubq}}$ isotopes are as follows.
	The Dirac Woods-Saxon basis is constructed in a box of $R_{box}$ = 20 fm, with the mesh of $\Delta r$ = 0.1 fm.
	For the particle-hole
	channel, the relativistic density functional PC-PK1 ~\cite{Zhao2010PRC82} is
	adopted.
    For the particle-particle channel, the density-dependent zero-range pairing force in Eq.~(\ref{eq:4}) is used, where the pairing strength $V_0 = -300.0 \ \text{MeV} \ \text{fm}^3$. The energy cutoff $E_{\text{cut}}^{+} = 300 \ \text{MeV}$ and the angular momentum cutoff $J_{\text{max}} = (31/2)\hbar$ are taken for the Dirac Woods-Saxon basis. The Legendre expansion truncation $\lambda_{\text{max}}$ for potentials and densities is set to 12.  
	The convergence has been checked in detail in Refs.~\cite{Zhang2020PRC102,Pan2022PRC106,Wang2024P7,Zhuang2026CPC}.

	\section{Results and discussion}
	\label{sec:4}

\subsection{Drip line and bulk properties}	
	
	
	\begin{figure}[htbp]
		\centering
		\includegraphics[scale=.35]{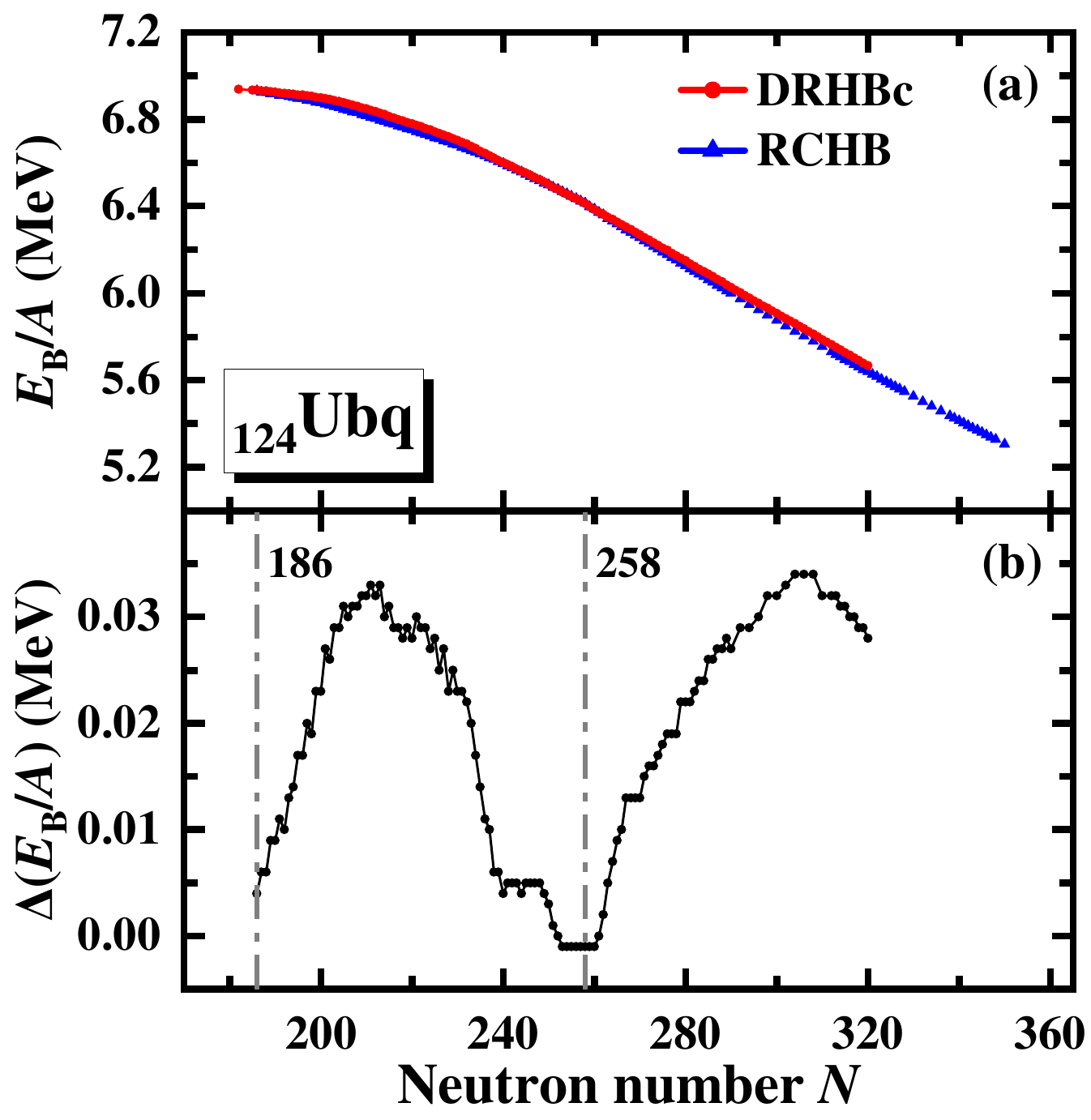}
		\caption{(Color online) (a) Binding energy per nucleon as a function of the neutron number for $\prescript{ }{124}{\text{Ubq}}$ isotopes in the DRHBc calculations. The results from the RCHB calculations are shown for comparison~\cite{Pan}.
			(b) Differences in binding energy per nucleon between the DRHBc and RCHB calculations, $\Delta(E_B/A)= (E_B/A)_{\mathrm{DRHBc}} - (E_B/A)_{\mathrm{RCHB}}$.}
		\label{fig:example1}
	\end{figure}
	
	Figure~$\ref{fig:example1}$(a) shows how binding energy per nucleon $E_{\mathrm B}/A$ changes with neutron number in the DRHBc calculations. The results from the RCHB calculations are also included for comparison~\cite{Pan}, serving as a spherical reference to investigate the deformation effects.
	As can be seen from the results of DRHBc, 
	the curve shows a monotonically decreasing trend, and the maximum of $E_{\mathrm B}/A$ occurs at $N=182$.
	This behavior can be attributed to the increasing asymmetry energy associated with the growing neutron number.
	Although the enlargement of the proton distribution slightly reduces the Coulomb repulsion and thereby tends to enhance the binding energy, this effect is insufficient to offset the increasing contribution from the asymmetry energy.

	Differences in binding energy per nucleon between the DRHBc and RCHB calculations are displayed in Fig.~\ref{fig:example1}(b).
	$\Delta(E_B/A)$ is nearly zero around $N = 184$ and $N = 258$.
	It first increases and then decreases up to $N = 258$, and rises again for $N > 258$,
	suggesting that $N = 184$ and $N = 258$ are possible shell closures.
	It is noted that $\Delta(E_B/A)$ remains within 35 keV. Although this value is small, for superheavy nuclei with $A > 300$, it corresponds to a difference in the total binding energy $E_{\mathrm{B}}$ of about 10.5 MeV, which is significant and highlights the important role of deformation effects.

	
	\begin{figure}[htbp]
		\centering
		\includegraphics[scale=.35]{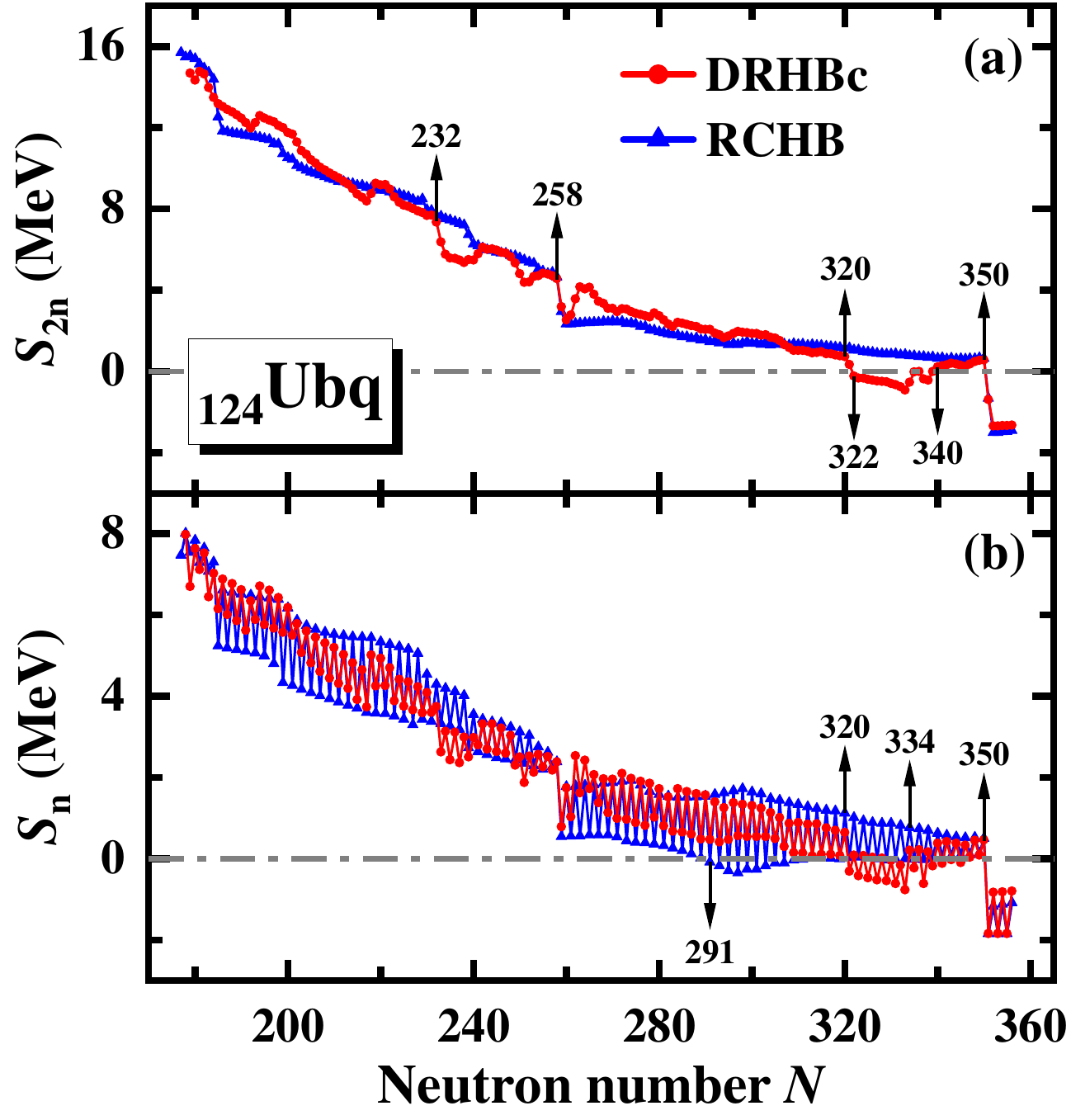}
		\caption{(Color online) Two-neutron separation energy (a) and one-neutron separation energy (b) as functions of the neutron number for $\prescript{ }{124}{\text{Ubq}}$ isotopes in the DRHBc calculations. The results from the RCHB calculations are shown for comparison. } 
		\label{fig:example2}
	\end{figure}
	
	The two-neutron and one-neutron separation energies can be calculated from the binding energy, thereby determining the position of the neutron drip line.
	Figure~$\ref{fig:example2}$ illustrates the two-neutron separation energy $S_{2n}$ and one-neutron separation energy $S_n$ of $\prescript{ }{124}{\text{Ubq}}$ isotopes obtained from the DRHBc calculations, in comparison with the RCHB results.
	For the DRHBc calculations, the two-neutron separation energy $S_{2n}$ becomes negative at $N = 322$, turns positive again at $N = 340$, and becomes negative once more at $N = 351$. 
	The one-neutron separation energy $S_n$ becomes negative starting from $N = 321$. In the ranges $N = 321$--$328$ and $N = 333$--$346$, an odd-even staggering occurs with negative values for odd $N$ and positive values for even $N$. $S_n$ is negative for $N = 329$--$332$ and positive for $N = 347$--$350$, and remains negative beyond $N = 351$.
	Negative values of $S_{n}$ and $S_{2n}$ indicate that these nuclei are unbound. 
	In comparison, for the results from the RCHB calculations, $S_{2n}$ remains positive up to $N = 350$, and becomes negative thereafter.
	 For $S_{n}$, in the ranges $N = 291$--$312$ and $N = 329$--$338$, an odd-even staggering occurs with negative values for odd $N$ and positive values for even $N$. $S_{n}$ remains positive for $N = 313$--$328$ and $N = 339$--$350$, and becomes negative again beyond $N = 351$.
	
	In Fig.~\ref{fig:example2}(a), 
	$S_{2n}$ evolves smoothly in the DRHBc calculations, but several abrupt changes occur at $N = 217, 232, 240, 252, 258, 260, 320,$ and $350$.
	In Fig.~\ref{fig:example2}(b), pronounced odd-even staggering is observed in the evolution of $S_{n}$, arising from pairing correlations. Moreover, at the neutron numbers where $S_{2n}$ exhibits abrupt changes, $S_{n}$ also undergoes changes that are more pronounced than the odd-even staggering.
	These abrupt changes may be related to deformation and shell effects. Among them, the sudden drops at $N = 258$ and $N = 350$ correspond to possible neutron shell closures, which are consistent with the results predicted by RCHB~\cite{Zhang2005NPA753,Liu2024P7,Pan2025P8}. However, from the results of DRHBc, there are two additional abrupt drops at $N = 232$ and $N = 320$, which may correspond to possible neutron subshell closures.
	Besides, the sharp increases at $N = 217$, 240, 252, and 260 stem from abrupt quadrupole deformation transitions.

	
	\begin{figure}[htbp]
		\centering
		\includegraphics[scale=.35]{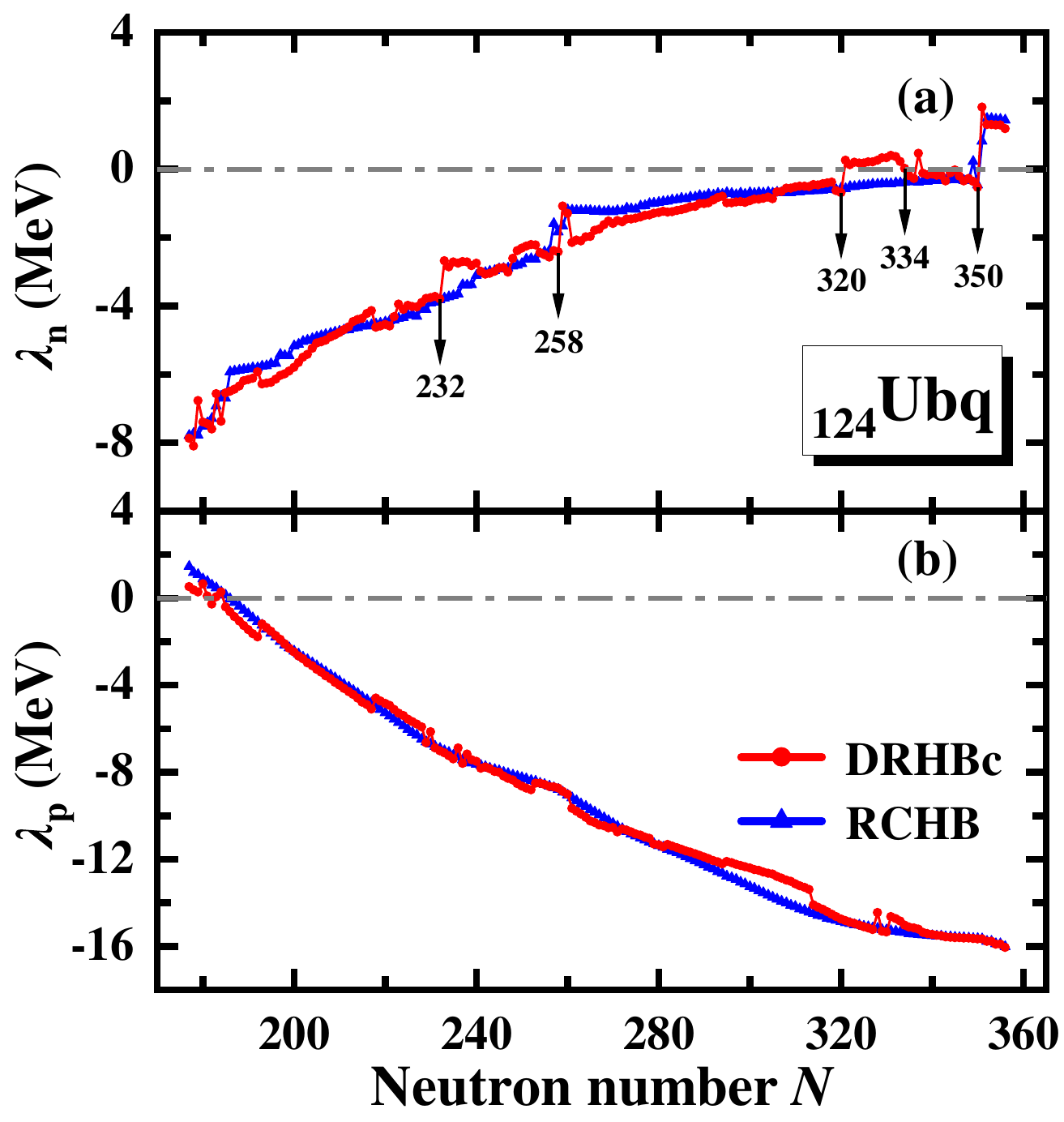}
		\caption{(Color online) Neutron (a) and proton (b) Fermi energies versus the neutron number for $\prescript{ }{124}{\text{Ubq}}$ isotopes in the DRHBc calculations. The results from the RCHB calculations are shown for comparison.} 
		\label{fig:example3}
	\end{figure}
	
\begin{figure}[htbp]
		\centering
		\includegraphics[scale=.4]{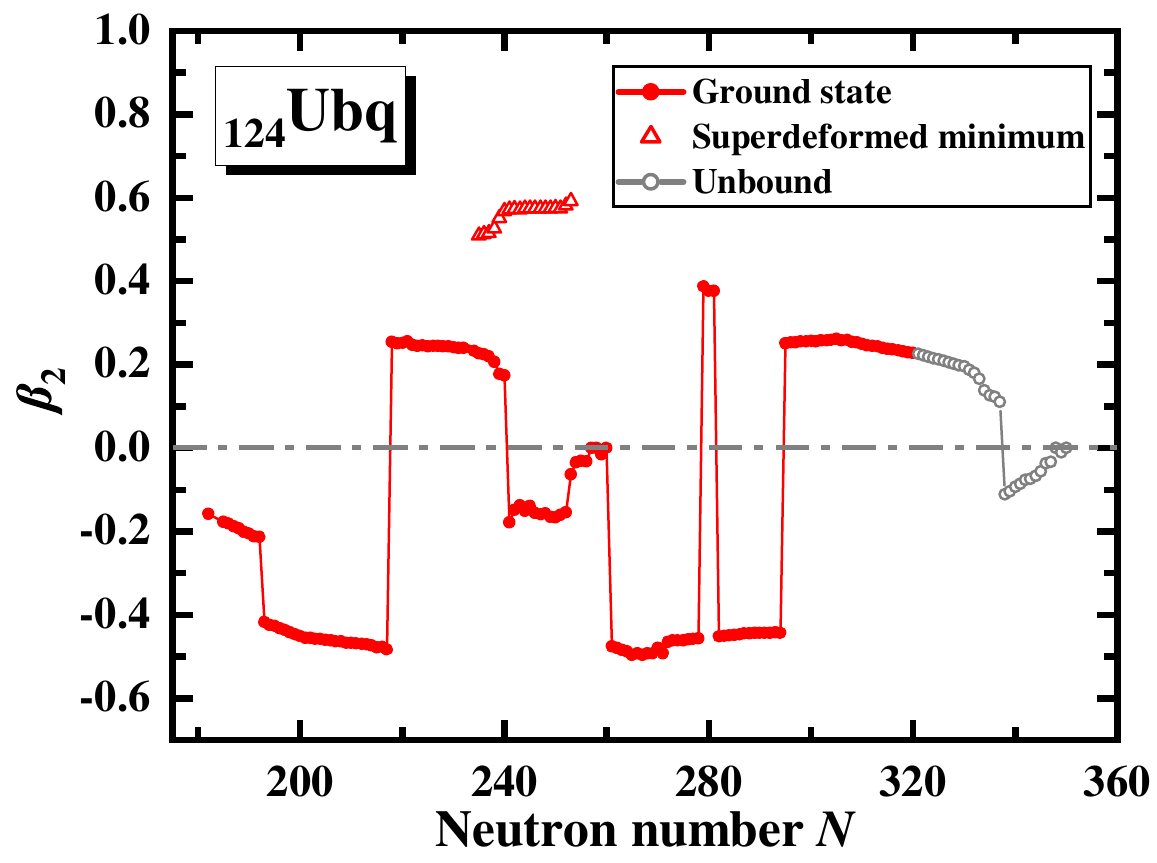}
		\caption{(Color online) Quadrupole deformation as a function of the neutron number for $\prescript{ }{124}{\text{Ubq}}$ isotopes in the DRHBc calculations.
		Red solid circles denote the selected ground states, red open
		triangles indicate calculated local minima around $\beta \approx 0.6$ that are not considered as ground states, and open gray circles indicate
		unbound nuclei.
	} 
		\label{fig:example5}
		
	\end{figure}

	In addition to the two-neutron and one-neutron separation energies, the neutron Fermi energy can also provide information about the neutron drip line, 
	as it represents the change of the total energy against the particle number in a mean-field level. Neutron (a) and proton (b) Fermi energies versus the neutron number for  $\prescript{}{124}{\text{Ubq}}$ isotopes in the DRHBc calculations are demonstrated in Fig.~\ref{fig:example3}, compared with the results from the RCHB calculations. 
	In Fig.~\ref{fig:example3}(a), from the results of DRHBc, there are sudden jumps in the neutron Fermi energy around $N = 258$ and $N = 350$, which correspond to possible neutron shell closures. Similarly, abrupt changes at $N = 232$ and $N = 320$ correspond to possible subshell closures.
    As the neutron number increases, the neutron Fermi energy remains negative for $N \leq 320$, becomes positive for $321 \leq N \leq 334$, turns negative again for $335 \leq N \leq 350$ except at $N = 337$, and becomes positive for $N \geq 351$.
    The increase in $\lambda_n$ at $N=337$ is associated with the occupation of the orbital in continuum, whereas the decrease in $\lambda_n$ beyond $N=337$ arises from changes in the deformed shell structure induced by the deformation change, as will be illustrated in Fig.~\ref{fig:example5}.
	In contrast, $\lambda_{n}$ predicted by the RCHB calculations maintains negative until $N = 350$, suggesting that the inclusion of deformation degrees of freedom affects the position of the neutron drip line.

	
	\begin{figure}[htbp]
		\centering
		\includegraphics[scale=.4]{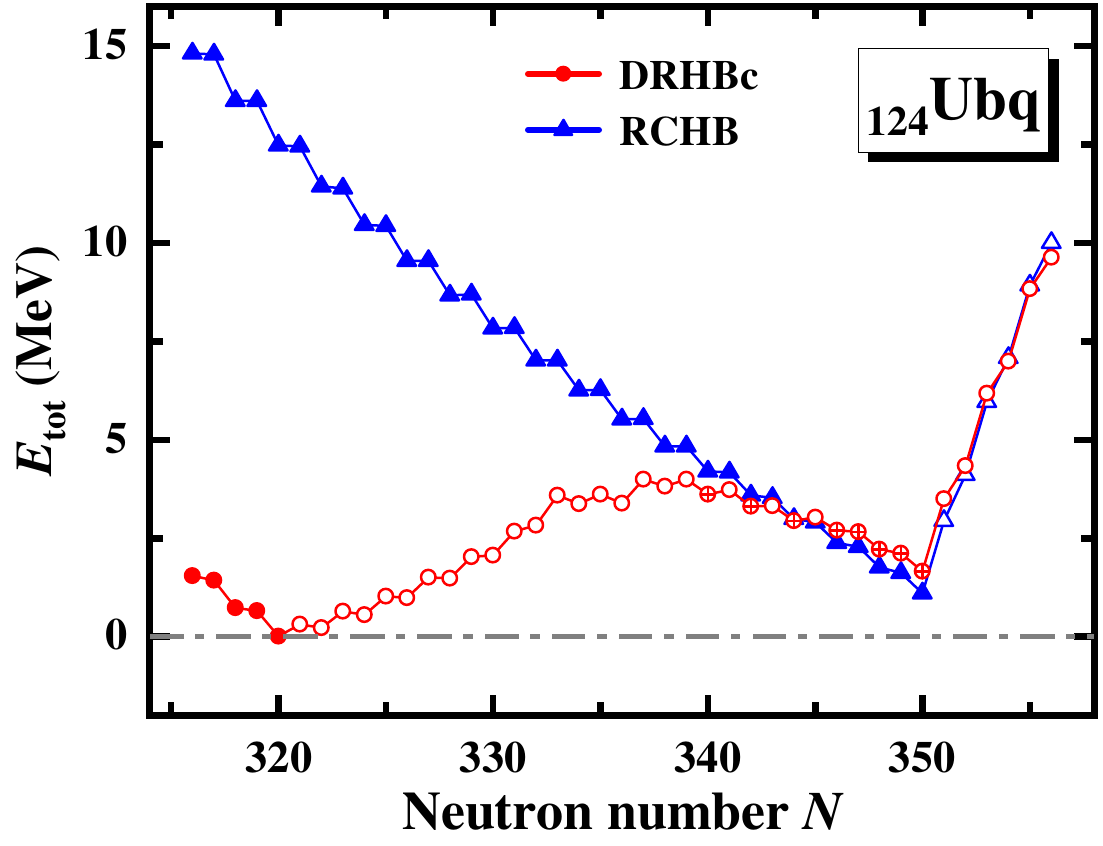}
		\caption{(Color online) Binding energy $E_{\text{B}}$ of $\prescript{ }{124}{\text{Ubq}}$ isotopes as a function of the neutron number near the neutron drip line in the DRHBc calculations. Solid, open, and crossed symbols denote bound nuclei, unbound nuclei, and multi-neutron emitters, respectively. The results from the RCHB calculations are shown for comparison. The binding energy of the dripline nucleus predicted by DRHBc ($\prescript{444 }{124}{\text{Ubq}}$) is normalized to zero, with the binding energies of other nuclei adjusted accordingly.	
		} 
		\label{fig:example4}
	\end{figure}

	In order to determine the dripline location of $\prescript{ }{124}{\text{Ubq}}$, Fig.~\ref{fig:example4} displays the binding energy $E_{\text{B}}$
	of $\prescript{ }{124}{\text{Ubq}}$ isotopes near the neutron drip line as a function of the neutron number in the DRHBc calculations, compared with the results from the RCHB calculations. 
	Although the two-neutron separation energies, one-neutron separation energies, and Fermi energies suggest that several nuclei beyond $N = 320$ remain bound, further verification is still required.
	The multi-neutron separation energy $S_{xn}$ is therefore examined. It is defined as	
	\begin{equation}
S_{xn}(Z,N)=B(Z,N)-B(Z,N-x),
	\end{equation}
	where $B(Z,N)$ denotes the binding energy of the nucleus with proton number $Z$ and neutron number $N$.
	Notably, it can be seen from the results of DRHBc that before $N = 320$ the binding energy rises with increasing neutron number, maintaining a positive separation energy. However, the binding energies of $\prescript{464}{124}{\text{Ubq}}$, $\prescript{466}{124}{\text{Ubq}}$, $\prescript{468}{124}{\text{Ubq}}$, and $\prescript{470-474}{124}{\text{Ubq}}$ are lower than that of $\prescript{444}{124}{\mathrm{Ubq}}$. 
	Consequently, 
	these isotopes are stable against two-neutron emission but unstable against multi-neutron emission due to positive two-neutron but negative multi-neutron separation energies.
	These nuclei are therefore identified as possible multi-neutron emitters. 
	The other nuclei have negative separation energies,  indicating that they are unbound. The so-called stability peninsula previously identified in the $50 \le Z \le 70$~\cite{Pan2021PRC104} and $101 \le Z \le 120$~\cite{Zhang2021PRC104,He2024PRC110,He2021CPC45} regions are not observed in our results.
	Based on the separation energy and the Fermi energy, all bound $\prescript{}{124}{\mathrm{Ubq}}$ isotopes are determined to be $\prescript{306}{124}{\text{Ubq}}$, $\prescript{309-444}{124}{\text{Ubq}}$. The last bound (dripline) nucleus is located at $N = 320$. 
	In contrast,
	the neutron drip line predicted by RCHB is located at
	$N = 350$, and no multi-neutron emitter beyond the drip line is found.
	The differences between the DRHBc and RCHB results regarding the neutron dripline location and the emergence of multi-neutron emitters highlight the impact of deformation effects~\cite{In2021IJMPE30}.

	
	Figure~$\ref{fig:example5}$ depicts the quadrupole deformation as a function of the neutron number for $\prescript{ }{124}{\text{Ubq}}$ isotopes in the DRHBc calculations. The shape evolution is as follows: (i) from $N=182$ to $N=217$, the nuclei are oblate; (ii) from $N=218$ to $N=240$, prolate; (iii) from $N=241$ to $252$, oblate; (iv) from $N=253$ to $N=260$, approximately spherical; (v) from $N=261$ to $N=294$, oblate except for $N=279, 280,$ and $281$, which are prolate; and (vi) from $N=295$ to $N=320$, prolate.
	Only a few nuclei near the possible neutron shell closure $N = 258 $ exhibit spherical shapes, while all other nuclei are deformed, including the previously mentioned possible subshell nuclei.

	For $\prescript{ }{124}{\text{Ubq}}$ isotopes, prolate minima with $\beta_2 \approx 0.6$, which are lower in energy than the corresponding minima at $|\beta_2| \approx 0.2$, are obtained in the DRHBc calculations and are denoted by open triangles in Fig.~\ref{fig:example5}.
	For superheavy nuclei, it is unlikely that the ground state exhibits such large deformations.
	As shown in Refs.~\cite{Lu2012PRC85,Lu2014PRC89,Zhou2016PS91}, 
	the inclusion of additional deformation degrees of freedom may significantly affect the potential energy surfaces. Although higher-order axial deformations ($\beta_{4}$, $\beta_{6}$, $\beta_{8}$, \ldots) are included self-consistently in the present DRHBc calculations, only the quadrupole deformation $\beta_{2}$ is constrained, and reflection asymmetric ($\beta_{3}$ and $\beta_{5}$) as well as triaxial degrees of freedom are not considered. The inclusion of these deformation degrees of freedom may help clarify this issue.
	In Ref.~\cite{Zhuang2026CPC}, when additional deformation degrees of freedom were taken into account, the large prolate minima around $\beta \approx 0.6$ disappeared, whereas the large oblate minima around $\beta \approx -0.5$ remained, supporting the present ground-state assignment.

	
	\begin{figure}[htbp]
		\centering
		\includegraphics[scale=.36]{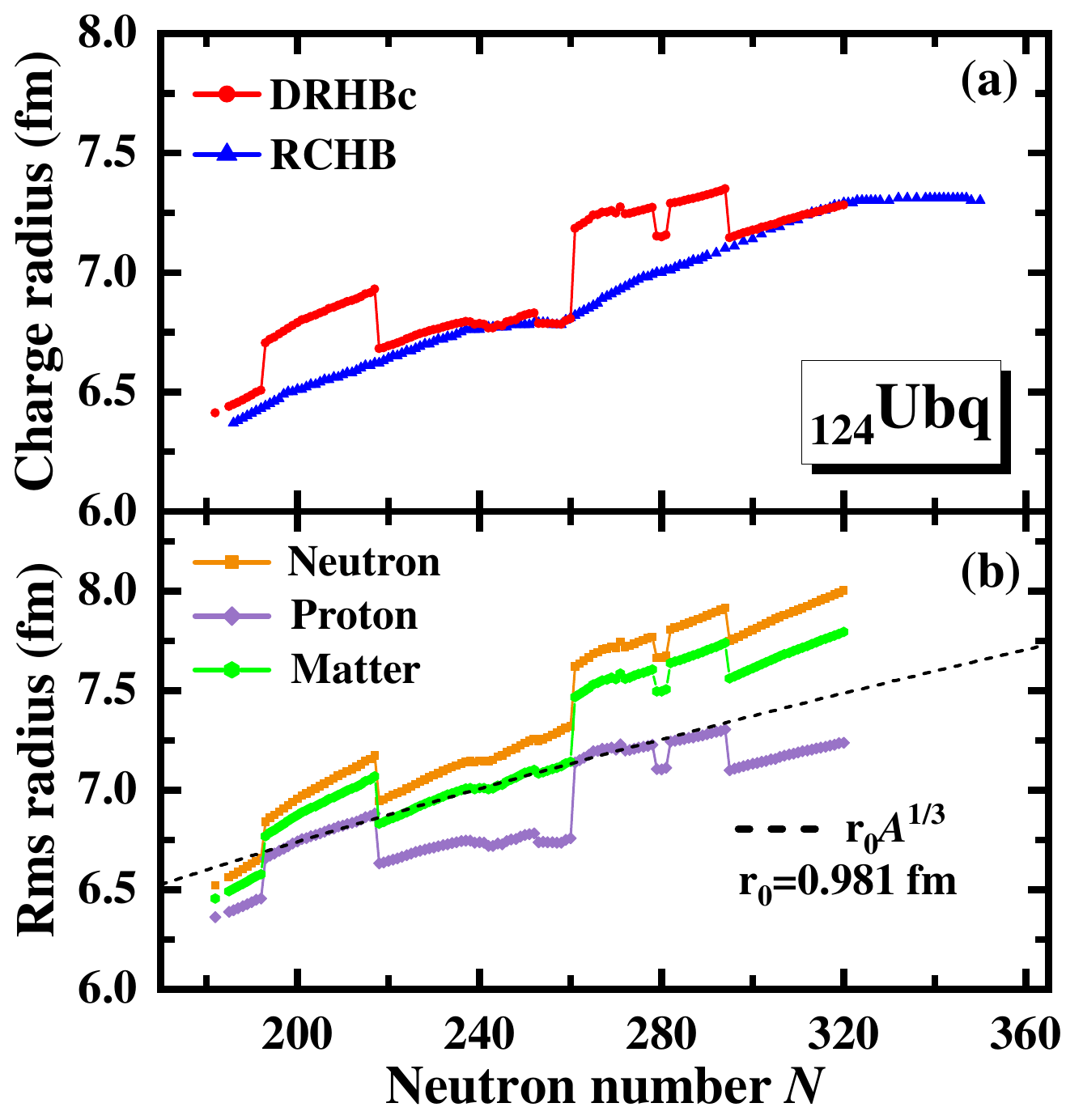}
		\caption{(Color online) Charge radius (a) and rms radius (b) as functions of the neutron number for $\prescript{ }{124}{\text{Ubq}}$ isotopes in the DRHBc calculations. Charge radius from the RCHB calculations is exhibited for comparison. The empirical matter radii $r_0 A^{1/3}$ are shown, in which $r_0 = R_m^\mathrm{DRHBc} / A^{1/3} = 0.981~\mathrm{fm}$
		determined by $\prescript{382}{124}{\text{Ubq}}$.
			} 
		\label{fig:example6}
		
	\end{figure}

     Figure~$\ref{fig:example6}$(a) presents the charge radius as a function of the neutron number for $\prescript{ }{124}{\text{Ubq}}$ isotopes in the DRHBc calculations, compared with the results from the RCHB calculations.
     The discrepancy between the DRHBc and RCHB calculations reveals the critical role of deformation effects in determining the bulk properties of superheavy nuclei~\cite{Pan2025PRC112}.
     
     In Fig.~\ref{fig:example6}(b), the root-mean-square (rms) neutron radii $R_n$, proton radii $R_p$, and matter radii $R_m$ calculated by DRHBc are shown together with the empirical matter radii $r_0 A^{1/3}$, where $r_0 = R_m^\mathrm{DRHBc} / A^{1/3} = 0.981~\mathrm{fm}$ is
     determined by $\prescript{382}{124}{\text{Ubq}}$. 
     As the neutron number increases, the deviations among $R_n$, $R_p$, and $R_m$ gradually expand, with $R_m$ situated between $R_n$ and $R_p$. For isotopes far from the valley of stability, the calculated $R_m$ values are systematically larger than the empirical estimates.

	\subsection{Shell structure}

	
	\begin{figure}[htbp]
		\centering
		\includegraphics[scale=.36]{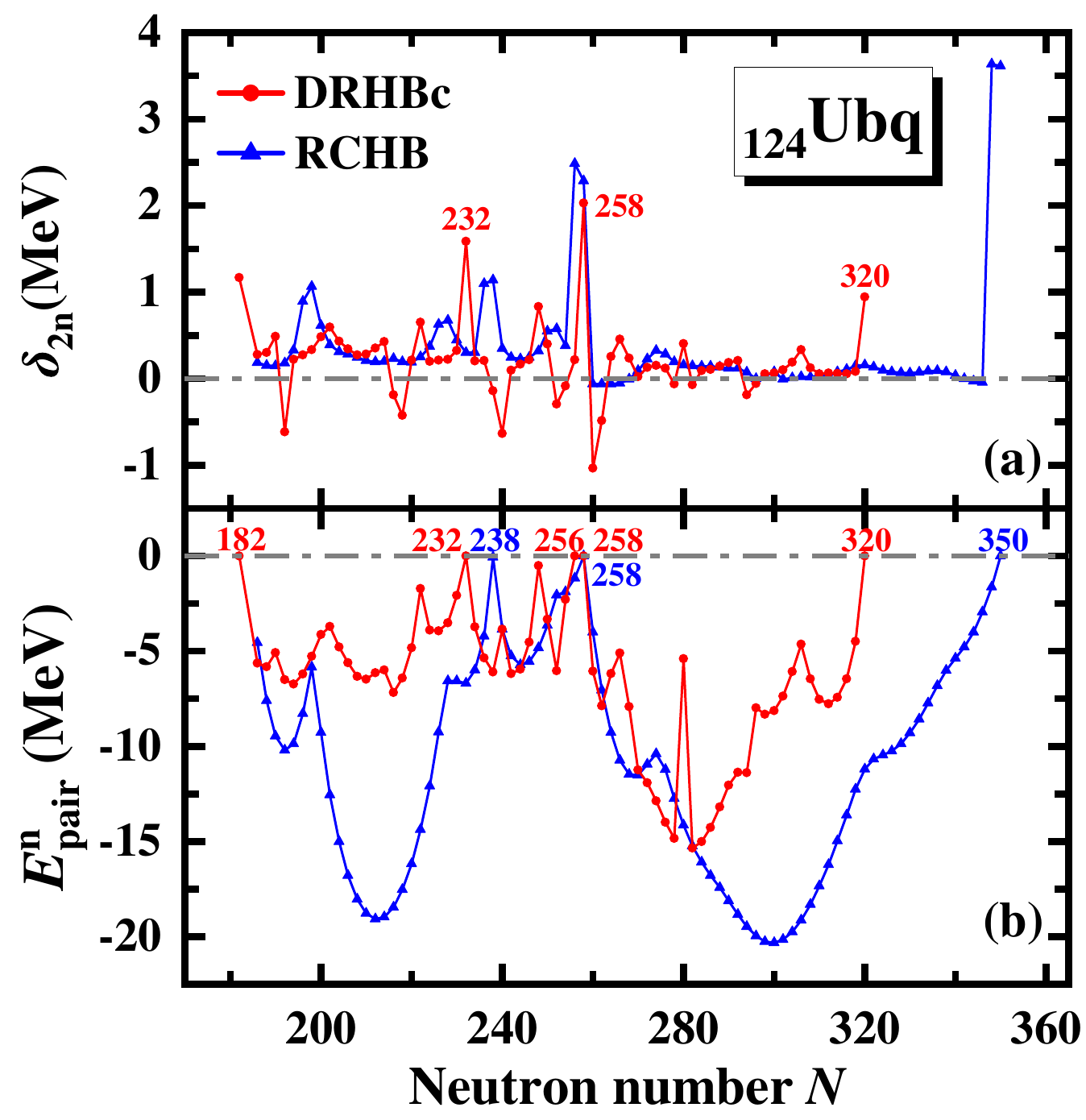}
		\caption{(Color online) Two-neutron gap $\delta_{2n}$ (a) and neutron pairing energies (b) as functions of the neutron number for even-even $\prescript{ }{124}{\text{Ubq}}$ isotopes in the DRHBc calculations. The results from the RCHB calculations are shown for comparison.} 
		\label{fig:example7}

	\end{figure}
	
    The previous discussion on neutron separation energy and neutron Fermi energy indicates the existence of potential neutron shell and subshell closures. 
    The two-neutron gap $\delta_{2n}$ is defined as $\delta_{\text{2n}}(Z, N) = S_{\text{2n}}(Z, N) - S_{\text{2n}}(Z, N + 2)$. 
    The peaks in $\delta_{2n}$, reflecting a drastic change in the two-neutron separation energy, serve as a signature of large shell gaps at the corresponding neutron numbers~\cite{Zhang2005NPA753,Li2014PLB732}.
    Compared with two-neutron separation energies, two-neutron gaps provide a more intuitive way to identify potential (sub)shell structures.
    The two-neutron gap as a function of the neutron number for even-even $\prescript{ }{124}{\text{Ubq}}$ isotopes in the DRHBc calculations is shown in Fig.~\ref{fig:example7}(a), compared with the results from the RCHB calculations. 
    The peaks of $\delta_{2n}$ can be observed at $N = 232$, 258, and 320, providing strong evidence for them as potential neutron shell and subshell closures. The other peaks originate from abrupt changes in deformation, as discussed above. In addition, the results from the RCHB calculations predict a peak at $N = 350$, suggesting an extra possible neutron shell closure.
    
    For nuclei with (sub)shell closures,
    density functional calculations give a pronounced gap in the single-particle spectrum, which prevents nucleons from scattering into higher orbitals and leads to the vanishing of pairing energy. Therefore, the sudden disappearance of pairing energy can also be used as a theoretical signature for the identification of possible (sub)shell closures. 
	 The neutron pairing energy as a function of the neutron number for even-even $\prescript{ }{124}{\text{Ubq}}$ isotopes in the DRHBc calculations is shown in Fig.~\ref{fig:example7}(b), compared with the results from the RCHB calculations. 
	The neutron pairing energy vanishes at $N = 232$, 258, and 320, providing evidence for the corresponding shell or subshell closures. Moreover, the results from the RCHB calculations show that the neutron pairing energy vanishes at $N = 350$, indicating an additional possible neutron shell closure.

	
	\begin{figure}[htbp]
		\centering
		\includegraphics[scale=.4]{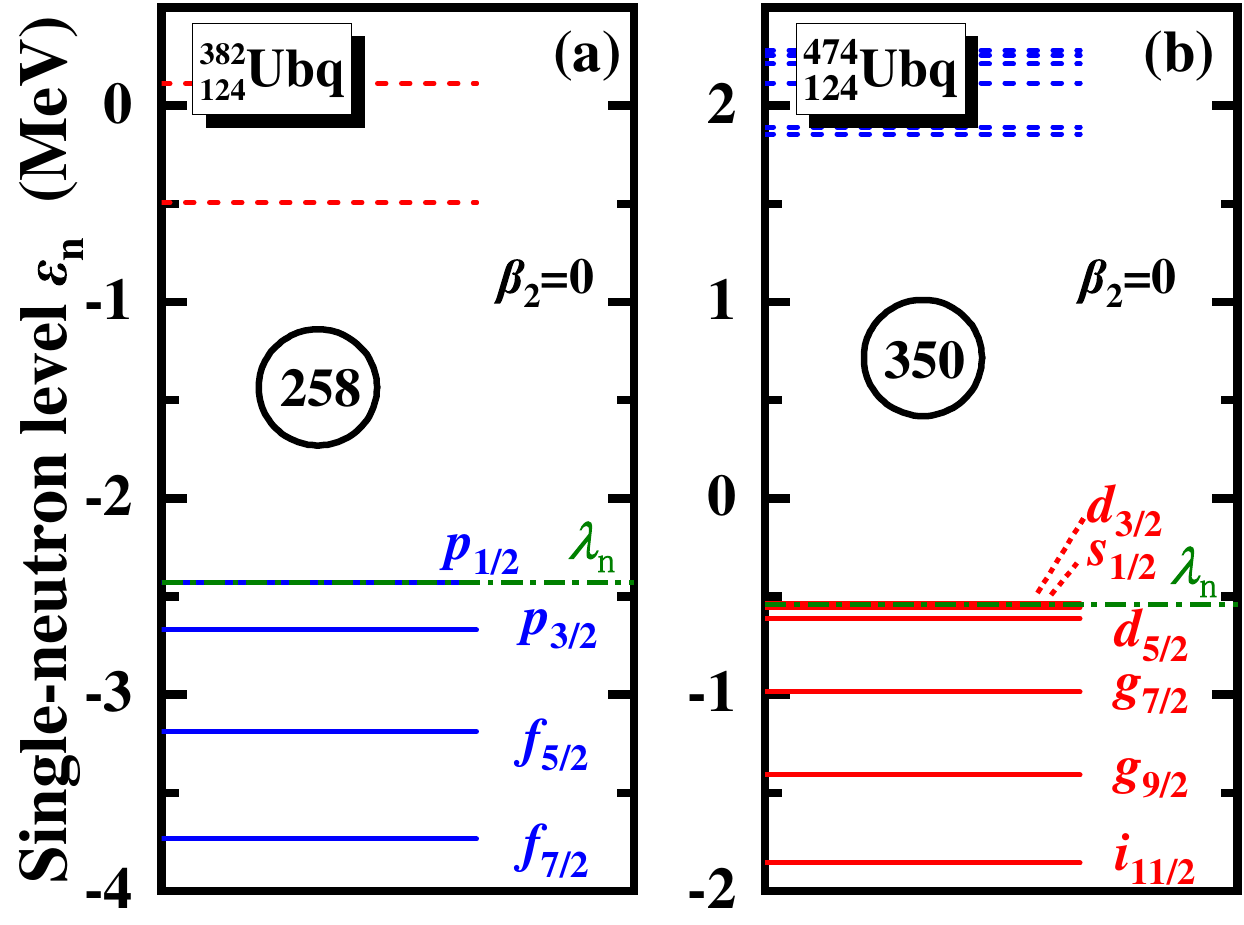}
		\caption{(Color online)  Single-neutron levels around the Fermi energy in the canonical basis for $\prescript{382}{124}{\text{Ubq}}$ and $\prescript{474}{124}{\text{Ubq}}$ in the DRHBc calculations. Solid and dashed lines indicate occupied and unoccupied orbitals, with red and blue representing positive and negative parity, respectively. The main components of the levels are given. Dot-dashed lines denote the neutron Fermi energy.} 
		\label{fig:example8}
	\end{figure}

      To further confirm $N = 258$ and 350 as potential neutron shell closures, single-neutron levels around the Fermi energy in the canonical basis for spherical $\prescript{382}{124}{\text{Ubq}}$ and $\prescript{474}{124}{\text{Ubq}}$ in the DRHBc calculations are illustrated in Fig.~\ref{fig:example8}. 
      Large gaps are observed above the neutron Fermi level of both nuclei, with gap sizes of approximately 1.93 and 2.39 MeV, respectively, providing strong evidence for
      $N = 258$ and 350 as potential neutron shell closures.

	
	\begin{figure}[htbp]
		\centering
		\includegraphics[scale=.4]{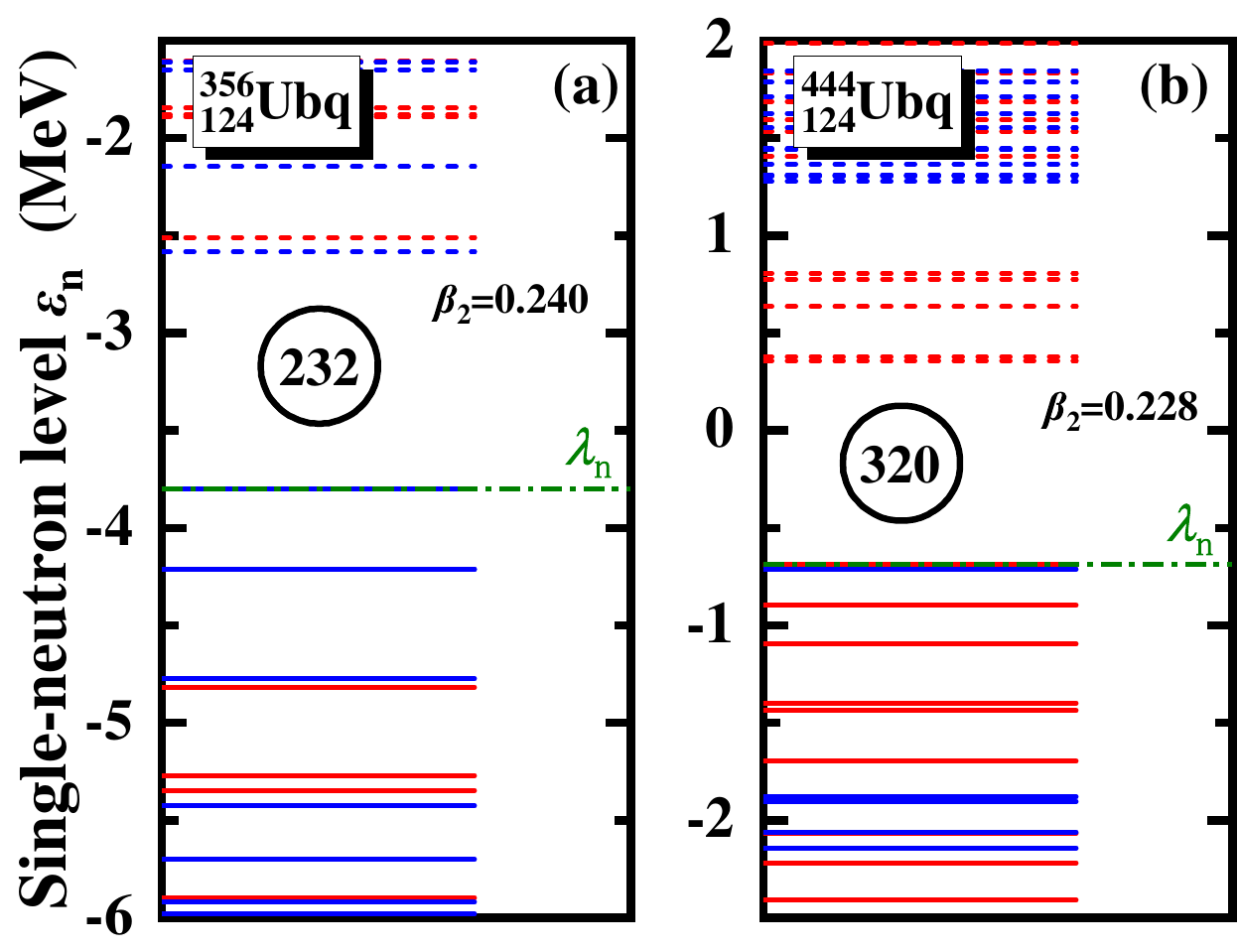}
		\caption{(Color online) Single-neutron levels around the Fermi energy in the canonical basis for $\prescript{356}{124}{\text{Ubq}}$ and $\prescript{444}{124}{\text{Ubq}}$ in the DRHBc calculations. Solid and dashed lines indicate occupied and unoccupied orbitals, with red and blue representing positive and negative parity, respectively. Dot-dashed lines denote the neutron Fermi energy. } 
		\label{fig:example9}
	\end{figure}

	In addition to the magic numbers associated with spherical nuclei, deformed subshell closures may also occur between magic numbers, as exemplified by the peaks of $\delta_{2n}$ at $N = 232$ and $N = 320$ in Fig.~\ref{fig:example7}(a). Figure~$\ref{fig:example9}$ presents the single-neutron levels around the Fermi energy in the canonical basis for deformed $\prescript{356}{124}{\text{Ubq}}$ and $\prescript{444}{124}{\text{Ubq}}$ in the DRHBc calculations.
	Due to deformation-induced level splitting, the number of single-particle levels shown here is significantly larger than that in Fig.~\ref{fig:example8}.
	Nevertheless, obvious gaps with magnitudes of approximately 1.22 and 1.04 MeV are observed for $\prescript{ }{124}{\text{Ubq}}$ isotopes with $N = 232$ and $320$, respectively, which strongly support them as possible neutron subshell closures.

	
	\begin{figure}[htbp]
		\centering
		\includegraphics[scale=.36]{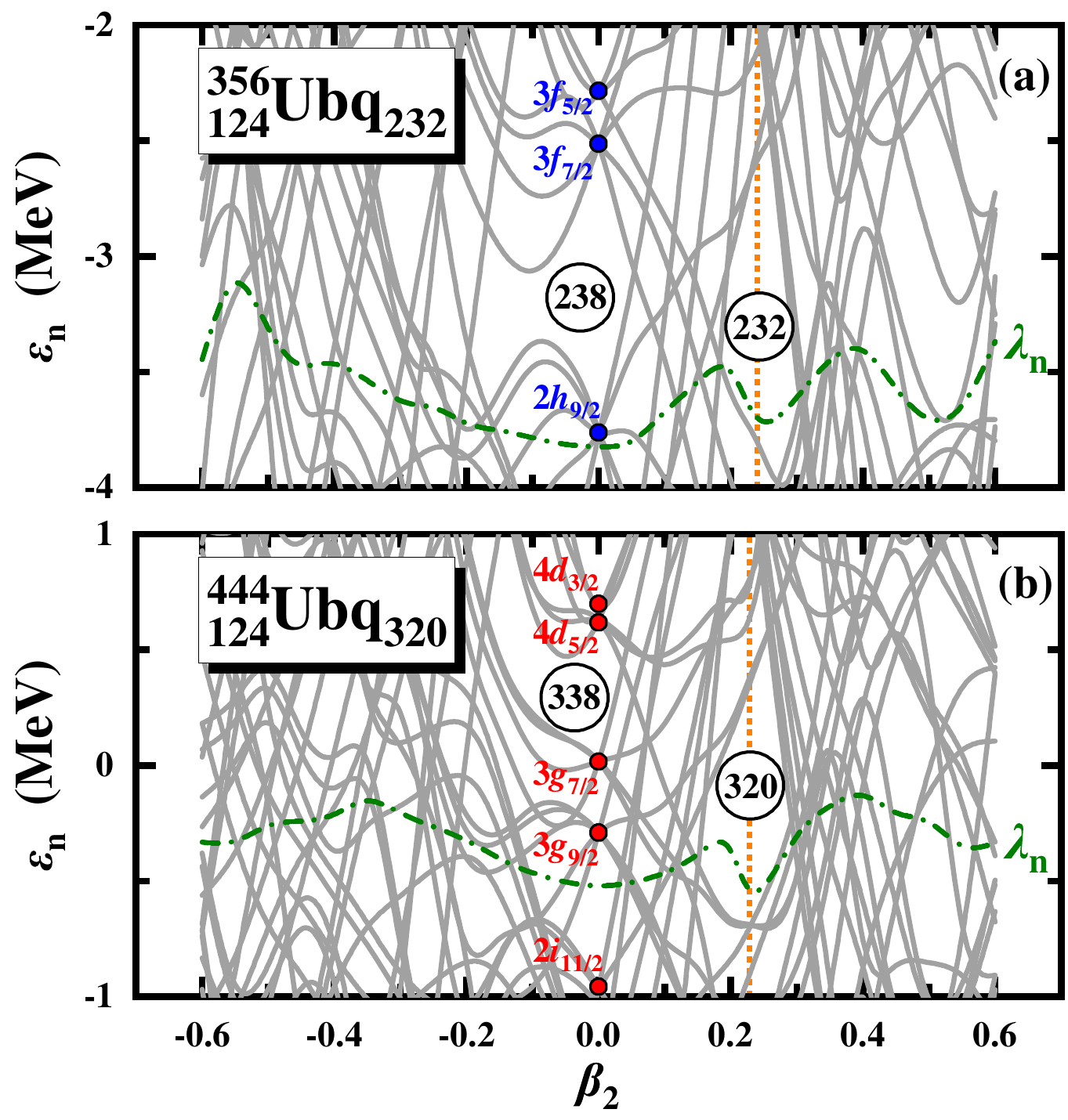}
		\caption{(Color online) Nilsson diagrams for neutrons in $\prescript{356}{124}{\text{Ubq}}$ (a) and $\prescript{444}{124}{\text{Ubq}}$ (b) as functions of quadrupole deformation. The neutron Fermi energies are represented with green dot-dashed lines. The ground-state deformations are shown by orange dotted lines. Spherical quantum numbers are marked by red and blue dots, corresponding to positive and negative parity, respectively.} 
		\label{fig:example10}

	\end{figure}
	
	 To explore the microscopic mechanism underlying the deformed subshells in Fig.~\ref{fig:example9}, the Nilsson diagrams for neutron orbitals in $\prescript{356}{124}{\text{Ubq}}$ and $\prescript{444}{124}{\text{Ubq}}$, as functions of quadrupole deformation $\beta_2$, are shown in Fig.~\ref{fig:example10}. 
	 For $\prescript{356}{124}{\text{Ubq}}$ in Fig.~\ref{fig:example10}(a), at $\beta_2=0$, a significant gap appears at $N = 238$,
	 which suggests that $N = 238$ is a possible spherical neutron subshell closure, consistent with the prediction in Ref.~\cite{Zhang2005NPA753}.
	  At the ground-state deformation $\beta_2=0.240$, the gap is located at $N=232$, resulting in a deformed subshell closure. 
	 For $\prescript{444}{124}{\text{Ubq}}$ in Fig.~\ref{fig:example10}(b), at $\beta_2=0$, a significant gap exists at $N = 338$, which has been predicted in previous work~\cite{Ismail2015JPGPP43}.
	 At the ground state with $\beta_2=0.228$, the gap is located at $N=320$, indicating a deformed subshell closure.

	
	\begin{figure}[htbp]
		\centering
		\includegraphics[scale=.45]{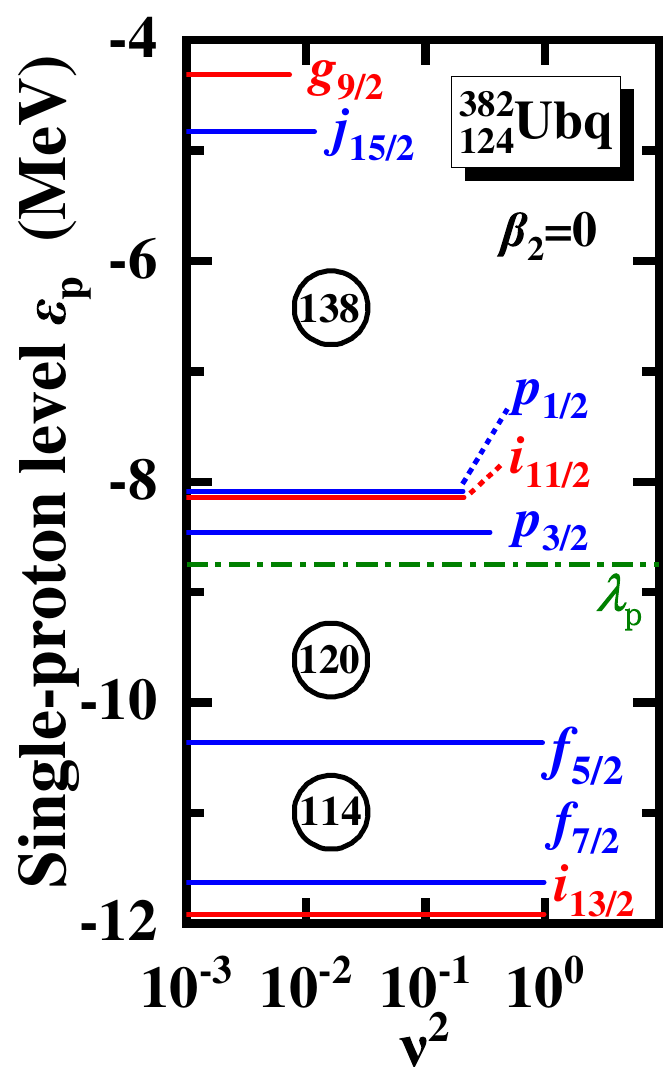}
		\caption{(Color online) Single-proton levels around the Fermi energy in the canonical basis versus the occupation probability $v^2$ for $\prescript{382}{124}{\text{Ubq}}$ in the DRHBc calculations. The main components of the levels are given. 
			Solid lines indicate occupied orbitals, with red and blue representing positive and negative parity, respectively.
			Dot-dashed lines denote the proton Fermi energy.} 
		\label{fig:example11}
	\end{figure}

  Previous studies have suggested the possibility of $Z = 124$ as a proton magic number through analyses based on the Strutinsky approach~\cite{Ismail2015JPGPP43} and the Skyrme mean-field approach~\cite{Asous2025IJMPE34}.
  Motivated by this, we have examined whether $Z = 124$ could be a potential proton shell closure.
     Single-proton levels around the Fermi energy in the canonical basis versus the occupation probability $v^2$ for $\prescript{382}{124}{\text{Ubq}}$ in the DRHBc calculations are depicted in Fig.~\ref{fig:example11}. 
     No significant gap is observed at $Z=124$, which indicates $Z = 124$ is not a robust proton magic number candidate in our calculations.
     In contrast, prominent gaps emerge at $Z = 114$, 120, and 138. In particular, the gap of 1.91 MeV at $Z = 120$ is larger than that of 1.26 MeV at $Z = 114$, consistent with the predictions of most relativistic energy density functionals (EDFs)~\cite{Bender1999PRC60,Zhang2024PRC110,Pan2025P8}. Therefore, $Z = 120$ and $Z = 138$ are suggested to be candidates for proton shell closures in the superheavy region within the present model, consistent with previous studies, such as Refs.~\cite{Zhang2005NPA753,Li2014PLB732}.

\subsection{Spontaneous decay}

	
	\begin{figure}[htbp]
		\centering
		\includegraphics[scale=.45]{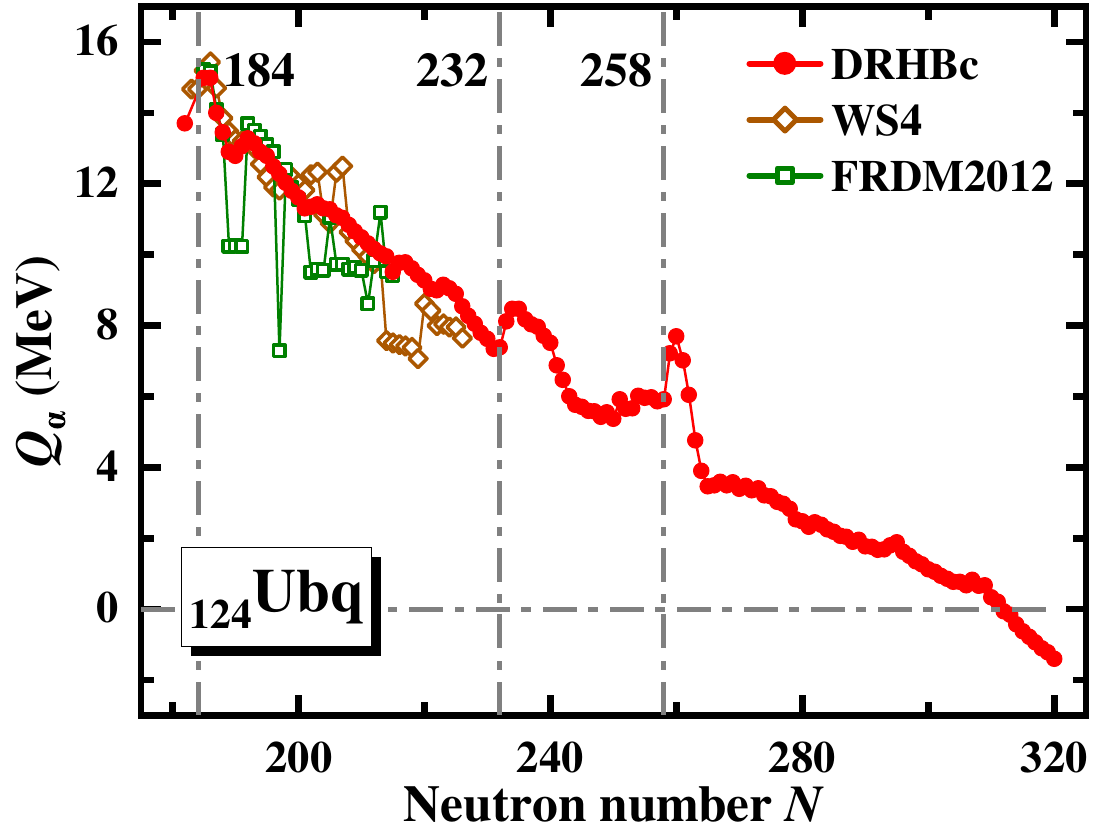}
		\caption{(Color online) The $Q_{\alpha}$ values of $\prescript{ }{124}{\text{Ubq}}$ as a function of the neutron number in the DRHBc calculations. The available results of WS4~\cite{Wang2014PLB734} and FRDM(2012)~\cite{Moeller2016ADNDT109}  mass models are displayed for comparison, where WS4 results are shown by green diamonds and FRDM(2012) results by brown squares.
		} 
		\label{fig:example12}
	\end{figure}

  The microscopic structures of Ubq isotopes discussed above are also expected to impact the stability of superheavy nuclei through their decay modes, such as $\alpha$ decay and spontaneous fission.

$\alpha$ decay is the principal decay channel of the heaviest nuclei synthesized to date, and is an essential tool for the identification and investigation of superheavy nuclei.
The $\alpha$ decay energy $Q_\alpha$, which can provide important information about their nuclear structure~\cite{Heenen2015NPA944}, is determined by
\begin{equation}
    Q_{\alpha}(A, Z) = B(A - 4, Z - 2) + B(4, 2) - B(A, Z).
    \end{equation}
	Figure~$\ref{fig:example12}$ presents the calculated $Q_{\alpha}$ values of $\prescript{ }{124}{\text{Ubq}}$ as a function of the neutron number from the DRHBc calculations, in comparison with available results from WS4 ~\cite{Wang2014PLB734} and FRDM(2012) ~\cite{Moeller2016ADNDT109} mass models. 
	The results of DRHBc reveal pronounced peaks immediately after $N = 232$ and $N = 258$, reflecting enhanced stability associated with corresponding subshell and shell closures, consistent with our preceding analysis.
	From the results of the WS4 mass model, a significant jump occurs at $N=184$, corresponding to a possible neutron shell closure. Results of DRHBc and FRDM(2012) also exhibit peaks near $N=184$. $\prescript{308}{124}{\text{Ubq}}$ is unbound in these two calculations and therefore omitted from the figure.
	All three calculations predict a gradual decrease of $Q_\alpha$ with increasing neutron 
	number, indicating reduced $\alpha$ decay energies and a weakening $\alpha$ decay tendency for heavier nuclei.

	
	\begin{figure}[htbp]
		\centering
		\includegraphics[scale=.45]{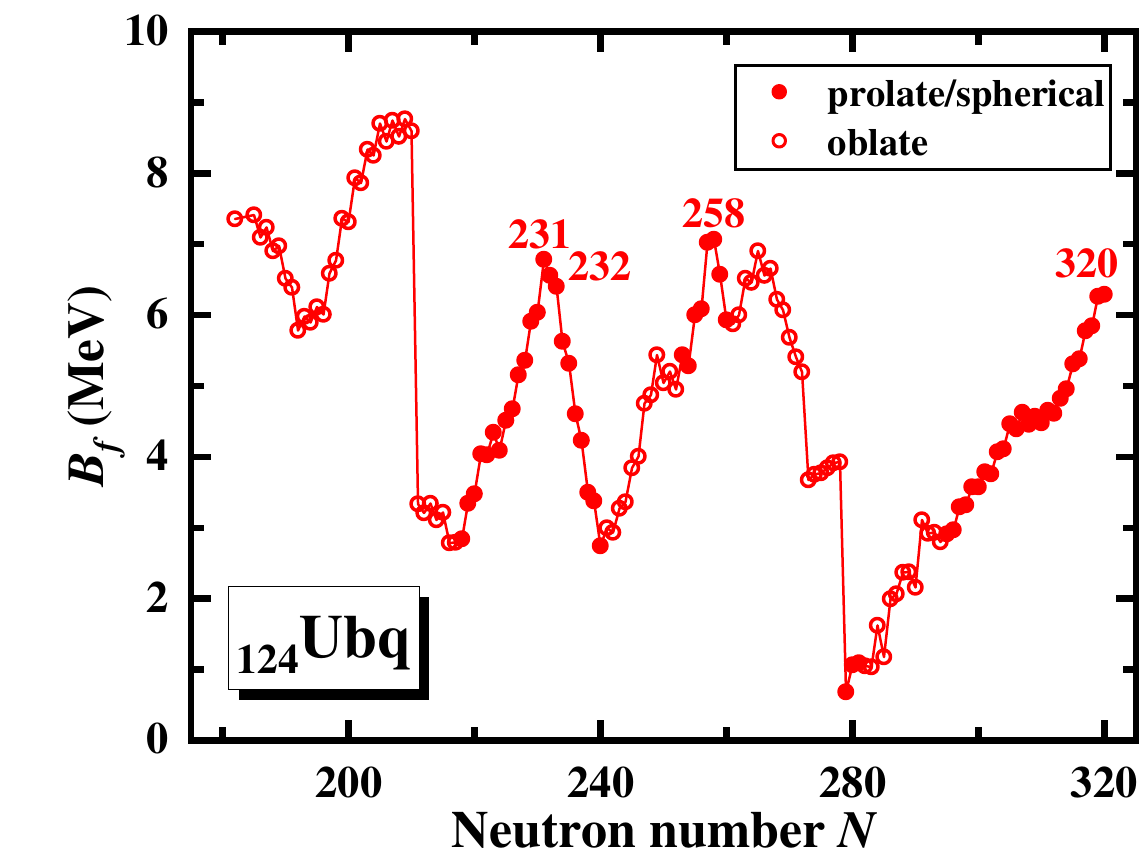}
		\caption{(Color online) Inner fission barrier heights of $\prescript{ }{124}{\text{Ubq}}$ in the DRHBc calculations. Solid circles indicate nuclei with prolate or spherical ground states, while open circles indicate nuclei with oblate ground states.
		} 
		\label{fig:example13}
	\end{figure}

	Spontaneous fission also plays an essential role in the synthesis of superheavy elements and the astrophysical $r$ process, with its probability closely related to the fission barrier.
	The fission barrier is essential for estimating fission half-lives, particularly the inner barrier, where a 1 MeV change in height may result in differences of several orders of magnitude.
	The inner fission barrier, defined as the energy difference between the ground state and the first barrier, exhibits systematic variations that reflect the influence of the underlying shell structure. 
	
	Since triaxial and octupole degrees of freedom, which are important for describing the inner fission barrier~\cite{Lu2012PRC85,Lu2014PRC89}, are not included in the present DRHBc theory,
	the potential energy surfaces are restricted to axial quadrupole deformation.
	Consequently, for many nuclei with oblate ground states, the calculated potential energy curves (PECs) exhibit a local maximum near $\beta_2 \approx 0$ when moving along the  path from the oblate ground-state minimum toward larger prolate deformation.
	However, since the spherical shape is not expected to represent a true fission saddle point, such a feature does not reflect the physical fission path, which would be modified by the inclusion of additional degrees of freedom. 
	Therefore, to avoid the unphysical ``spherical'' fission barrier, in this work, the inner fission barrier height is taken as the energy difference between the ground state and the first local maximum encountered along the prolate deformation path beyond $\beta_2 = 0.1$.
	For nuclei with oblate ground states, if no local maximum is found along the prolate deformation, the first barrier is searched for along increasing oblate deformation.

	Figure~\ref{fig:example13} illustrates the inner fission barrier heights $B_f$ of $\prescript{ }{124}{\text{Ubq}}$ in the DRHBc calculations. 
	It can be observed that prominent peaks appear at the possible neutron shell or subshell closures (e.g., $N = 232, 258$, and $320$). 
	The shell-driven enhancement of $B_f$ increases the stability of superheavy nuclei against SF. Conversely, in midshell regions, reduced shell stabilization leads to a sharp decrease in $B_f$. 
	
\begin{figure}[htbp]
	\centering
	\includegraphics[scale=.35]{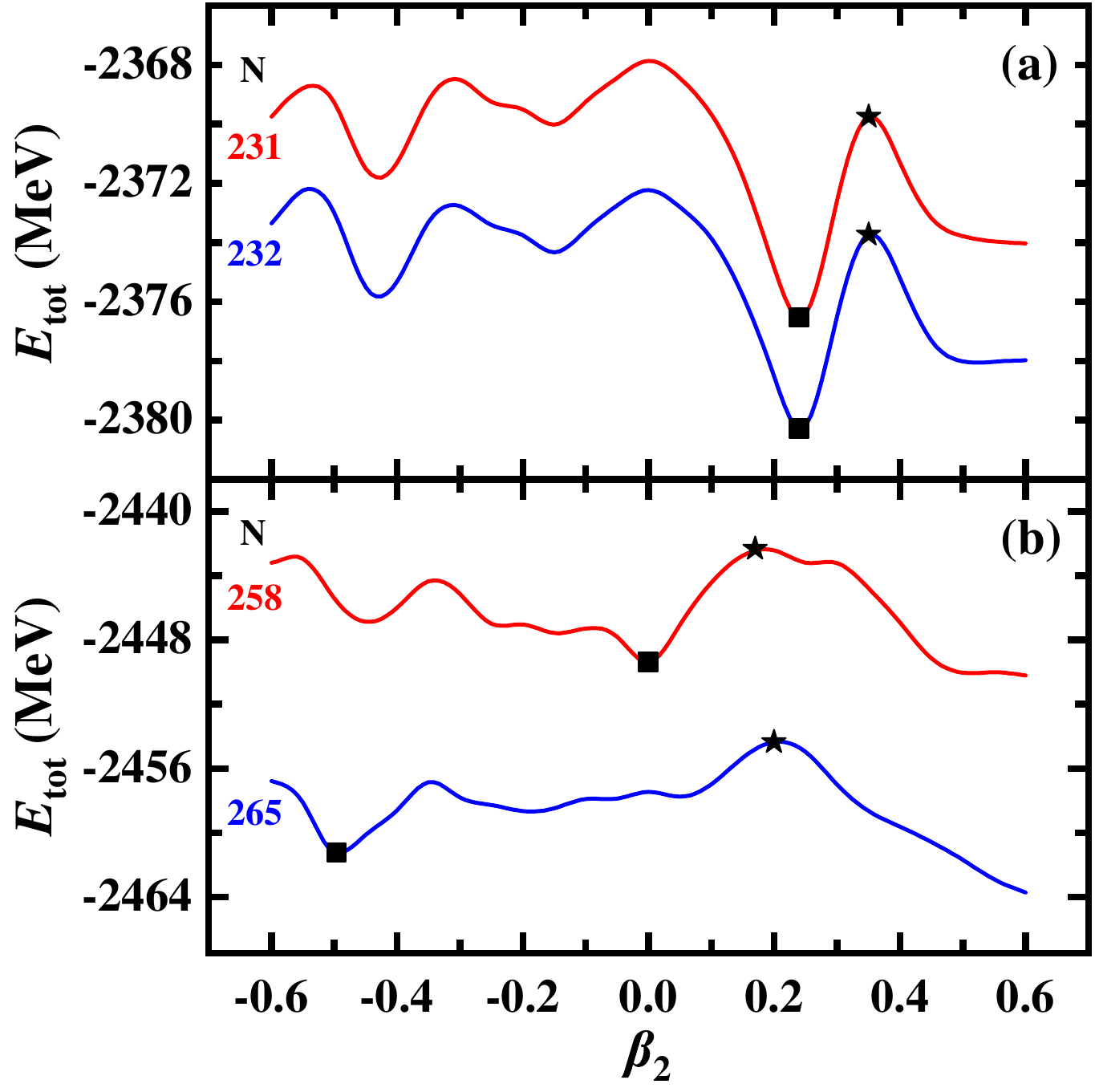}
	\caption{(Color online) 
		Potential energy curves of selected nuclei illustrating the two special cases in Fig.~\ref{fig:example13}: (a) $\prescript{355,356}{124}{\text{Ubq}}$; (b) $\prescript{382,389}{124}{\text{Ubq}}$.
		Squares represent the ground states, while stars indicate the maxima used to calculate the inner fission barrier.}
	\label{fig:example14}
\end{figure}

	As demonstrated in Fig.~\ref{fig:example13}, the inner fission barrier of $N=231$ is higher than that of $N=232$ by about 0.22 MeV. 
	To examine this feature, the PECs of $\prescript{355}{124}{\text{Ubq}}$ and $\prescript{356}{124}{\text{Ubq}}$ are displayed in Fig.~\ref{fig:example14}(a). It is found that the PECs of these two nuclei evolve similarly, resulting in only a small difference in the inner fission barrier.

	In addition, it is found in Fig.~\ref{fig:example13} that the inner fission barrier around $N=265$ is close to that at the magic number $N=258$. To further analyze this behavior, the PECs of $\prescript{382}{124}{\text{Ubq}}$ and $\prescript{389}{124}{\text{Ubq}}$ are illustrated in Fig.~\ref{fig:example14}(b). It is evident that both nuclei exhibit a local maximum around $\beta_2 = 0.2$. However, the ground state of $N=258$ is spherical, whereas for $N=265$ the energy of oblate minimum lies lower. As a result, the barrier increases at $N=265$.

	It should be emphasized that such behaviors of the PECs may be associated with triaxial deformation effects. 
	In particular, the triaxial extension of the DRHBc theory, the triaxial RHB theory in continuum (TRHBc), has been developed recently~\cite{Zhang2023PRC108,Zhang2025PRC112a}. Future investigations based on the TRHBc theory may help elucidate the effect of triaxial deformation on these nuclei.


\begin{figure}[htbp]
	\centering
	\includegraphics[scale=.39]{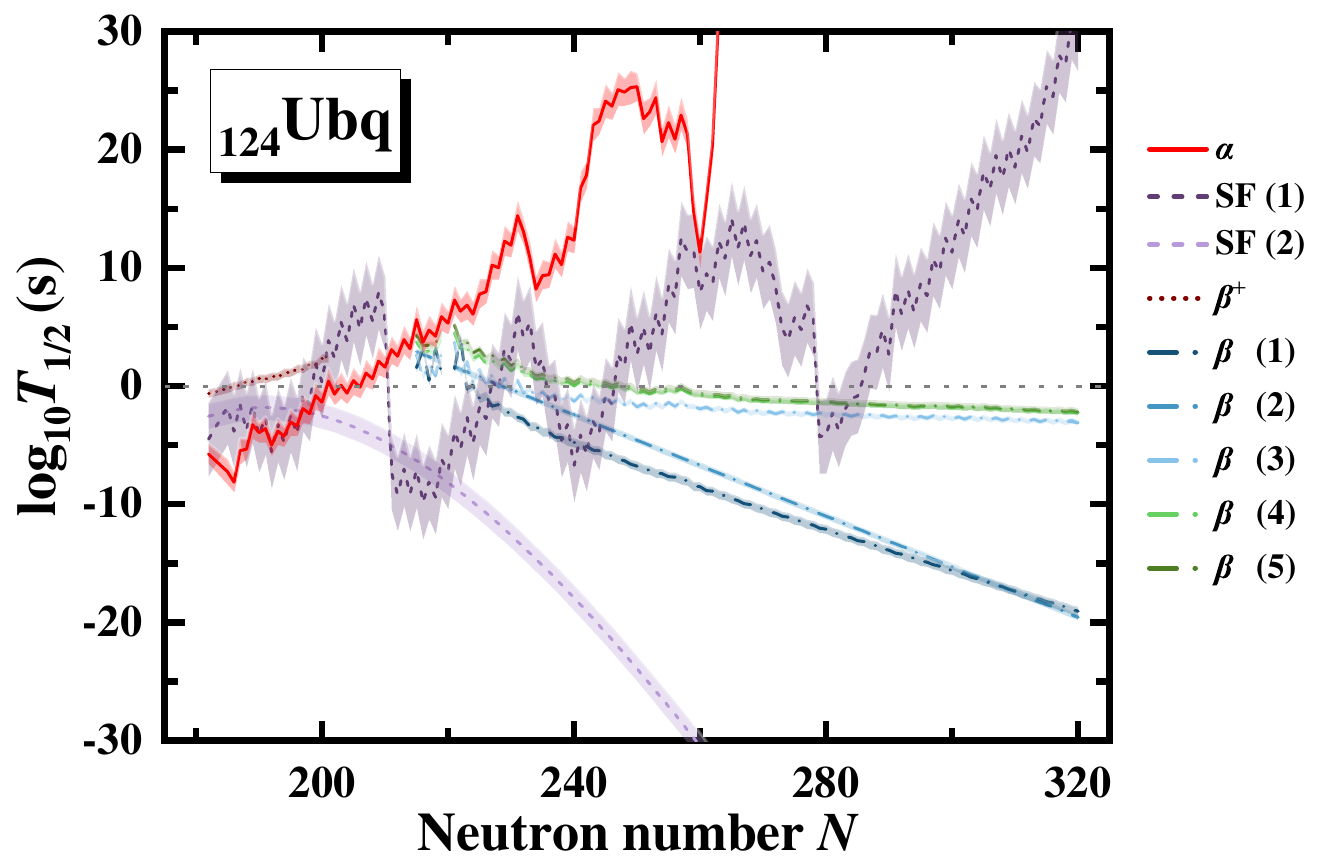}
	\caption{(Color online) Half-lives of $\prescript{ }{124}{\text{Ubq}}$ isotopes for $\alpha$ decay~\cite{Peng2009NPR26,Ismail2025CPC49}, SF~\cite{KARPOV2012IJMPE21,Pahlavani2023MPLA38}, $\beta^+$ decay~\cite{Sobhani2023CJP85}, and $\beta^-$ decay~\cite{Sobhani2023CJP85,Zhang2007JoPGNaPP34,Zhou2017SCMA60,JinGe2024APS73,Tian2025CPC49} predicted using different semiempirical formulas are represented by solid, short-dashed, dotted, and dot-dashed lines, respectively.
	The uncertainties estimated by the rms deviations between the logarithms of the calculated and experimental half-lives within the fitting ranges are indicated by shaded regions in the corresponding colors. See text for details. } 
	\label{fig:example15}
\end{figure}

In addition to $\alpha$ decay and SF, $\beta$ decay is also a typical decay mode in superheavy nuclei.
To clarify the dominant decay channels of the $Z=124$ isotopes, we have calculated the half-lives of $\prescript{ }{124}{\text{Ubq}}$ isotopes for $\alpha$ decay, SF, $\beta^+$ decay, and $\beta^-$ decay 
using different semiempirical formulas, together with the uncertainties estimated by the rms deviations between the logarithms of the calculated and experimental half-lives within the fitting ranges, as presented in Fig.~\ref{fig:example15}.
The $\alpha$ decay half-lives have been calculated using the Viola–Seaborg (VS) formula~\cite{Viola1966JINC28} with the ``New2'' parameter set provided in Ref.~\cite{Peng2009NPR26}, and using the NRDX formula~\cite{Ni2008PRC78}, universal decay law (UDL)~\cite{Qi2009PRL103}, and Royer formula~\cite{Royer2000JPGNPP26} with the parameter set given in Table~1 of Ref.~\cite{Ismail2025CPC49}.
The SF half-lives are computed using two formulas, denoted as SF(1) and SF(2), with the parameter sets taken from Refs.~\cite{KARPOV2012IJMPE21} and \cite{Pahlavani2023MPLA38}, respectively.
As for $\beta$ decay half-lives, 
we first employed a general formula of 
Eq.~(2.1) in Ref.~\cite{Sobhani2023CJP85} with the parameters listed in Table~1 therein, which is valid for both $\beta^+$ and $\beta^-$ decay, to calculate the half-lives shown as $\beta^+$ and $\beta^-$(1) in Fig.~\ref{fig:example15}.
In addition, four other semiempirical formulas for $\beta^-$ decay half-lives are also employed, taken from Eqs.~(17), (9), (8), and (9) of Refs.~\cite{Zhang2007JoPGNaPP34,Zhou2017SCMA60,JinGe2024APS73,Tian2025CPC49}, with the results labeled as $\beta^-$(2), (3), (4), (5) in Fig.~\ref{fig:example15}, respectively.

Since the parameters of the empirical formulas were determined by fitting to experimental data, their applicability and reliability also depend on the region of the nuclear chart covered by the fitting data. The Appendix therefore summarizes the fitting ranges and deviations reported in the original references or calculated by us, together with their explicit forms and corresponding parameters.

For $\alpha$ decay, 
in Fig.~\ref{fig:example15}, the half-life is taken as the average of the predictions from the four formulas, and the shaded region represents the union of their respective deviation ranges. The relatively narrow shaded region, with a width of less than 1.6, indicates that the four formulas give consistent predictions. This consistency can be attributed to that semiempirical formulas for $\alpha$ decay half-life are well established.

$\beta^+$ decay is predicted to occur only in nuclei with $N \le 203$, except for $N = 202$, and its half-life is longer than those of both $\alpha$ decay and SF. $\beta^-$ decay may occur only for $N \ge 213$, except for $N = 214$ and $N=220$. For $\beta^-$ decay, these semiempirical formulas show a consistent trend prediction: the $\beta^-$ decay half-lives decrease with increasing neutron number. However, noticeable differences remain in their quantitative predictions. In particular, formulas $\beta^-(1)$ and $\beta^-(2)$ tend to predict a more rapid decrease in half-life, whereas $\beta^-(3)$, $\beta^-(4)$, and $\beta^-(5)$ give relatively similar results.
Specifically, $\beta^-(1)$ requires $Q$ values of the related branch of beta decay and the parent nucleus. $\beta^-(2)$ is based on an exponential dependence of the half-life on nucleon numbers $(Z, N)$ for nuclei far from the $\beta$-stability line, and further incorporates shell effects. $\beta^-(3)$ establishes a relationship between the half-life and the $Q$ value, nucleon numbers $(Z, N)$, and neutron excess $(N - Z)/A$, with additional corrections for pairing and shell effects. $\beta^-(4)$ takes into account the pairing effect, the shell effect, and the isospin dependence. $\beta^-(5)$ introduces the transition-strength contribution to the $\beta$ decay half-life.
Overall, although these formulas exhibit consistent trends, their quantitative differences arise from distinct theoretical assumptions, fitting datasets, and treatments of nuclear structure effects, such as pairing and shell corrections.

The SF(1) formula explicitly incorporates the fission barrier height and therefore predicts enhanced SF half-lives around the candidate (sub)shell closures at \(N=232\), 258 and 320. In contrast, the SF(2) formula is fitted based on the fissility parameter $Z^2/A$ and the isospin parameter $(N - Z)/(N + Z)$, without accounting for shell effects in the superheavy region. Consequently, SF(2) fails to reproduce the variations of the fission barriers shown in Fig.~\ref{fig:example13}, and we take SF(1) as the SF half-lives in the following discussion.

For each isotope, the decay mode with the shortest predicted half-life is regarded as the dominant decay channel.
For $N<190$, $\alpha$ decay is the dominant decay mode. 
For $190 \le N \le 198$, the predicted $\alpha$ decay and SF half-lives are comparable, indicating a strong competition between these two decay modes. 
For $199 \le N \le 210$, $\alpha$ decay remains dominant.
For $210 < N < 227$, SF becomes the dominant decay mode.
For $N \ge 227$, $\beta^-$ decay is the dominant decay mode for most nuclei.

It should be emphasized that this study provides only an exploratory estimation of nuclear half-lives, based on ground-state properties obtained from the DRHBc theory, in combination with phenomenological formulas from Refs.~\cite{ Peng2009NPR26,Ismail2025CPC49,KARPOV2012IJMPE21,Pahlavani2023MPLA38,Sobhani2023CJP85,Zhang2007JoPGNaPP34,Zhou2017SCMA60,JinGe2024APS73,Tian2025CPC49}. 
Nevertheless, when the predicted half-lives of competing decay modes differ by several orders of magnitude, the qualitative determination of the dominant decay mode can still provide important reference. 
Future work is expected to improve the present framework by incorporating additional microscopic nuclear effects, which may lead to more reliable and precise predictions.

	\section{Summary}
	\label{sec:6}

	The ground-state properties of the $Z = 124$ isotopes have been systematically investigated using the DRHBc theory with the density functional PC-PK1.
	The neutron dripline location is determined based on separation energies and Fermi energies, and is located at $N = 320$. The stability of nuclei against multi-neutron emission is examined through multi-neutron separation energies, and eight possible multi-neutron emitters beyond the neutron drip line are identified. By comparing the results of DRHBc and RCHB, the impact of deformation effects on ground-state properties is clearly demonstrated.

	Furthermore, we have systematically explored the potential neutron shell closures at $N = 258$ and $N = 350$, as well as two possible neutron subshell closures at $N = 232$ and $N = 320$. 
	Besides, our results do not support $Z = 124$ as a robust proton magic number. In contrast, $Z = 120$ and $138$ are identified as more favorable candidates for proton shell closures.
    These results contribute further theoretical insights into the evolution of nuclear shell structures and the emergence of the ``island of stability'' for superheavy elements.
	
	Finally, the half-lives for $\alpha$ decay, $\beta$ decay, and SF
	of $Z = 124$ isotopes are estimated using various semiempirical formulas. 
	The dominant decay modes of the $^{124}$Ubq isotopes are predicted by comparing the half-lives of different decay channels.
	For $N<190$, $\alpha$ decay is the dominant decay mode. 
	For $190 \le N \le 198$, the predicted $\alpha$ decay and SF half-lives are comparable, indicating a strong competition between these two decay modes. 
	For $199 \le N \le 210$, $\alpha$ decay remains dominant.
	For $210 < N < 227$, SF becomes the dominant decay mode.
	For $N \ge 227$, $\beta^-$ decay is the dominant decay mode for most nuclei.

	These results provide a preliminary theoretical insight into the synthesis and stability of next-generation superheavy nuclei. Future improvements are expected to incorporate more nuclear effects, which may lead to more reliable and quantitatively accurate predictions.

\begin{acknowledgments}
	The authors would like to thank members of the DRHBc Mass Table Collaboration for their helpful discussions, in particular Professor Kaiyuan Zhang, Professor Shuangquan Zhang, Professor Wei Zhang and Professor Qiang Zhao for their valuable suggestions and guidance.
	This work was supported by National Natural Science Foundation of China under Grant No. 12605194 and the Fundamental Research Funds for the Central Universities (JUSRP202406002).
\end{acknowledgments}

\appendix
\renewcommand{\thesection}{}
\renewcommand{\thesubsection}{\arabic{subsection}}

\section{Semiempirical formulas for decay half-lives}
\label{app:decay}

\subsection{$\alpha$ decay}
The $\alpha$ decay half-lives  (in seconds) have been estimated using four formulas: the Viola-Seaborg (VS) formula~\cite{Viola1966JINC28}, the NRDX formula~\cite{Ni2008PRC78}, the universal decay law (UDL)~\cite{Qi2009PRL103}, and the Royer formula~\cite{Royer2000JPGNPP26}.

The VS formula is given by
{\setlength{\abovedisplayskip}{4pt}
	\setlength{\belowdisplayskip}{4pt}
	\begin{equation}
	\log_{10}T_{1/2}^{\rm VS}
	=(aZ+b)Q_{\alpha}^{-1/2}
	+(cZ+d)+h_{\log},
	\label{eq:A1}
	\end{equation}
where $Z$ is the proton number of the parent nucleus, $Q_{\alpha}$ (in MeV) is the $\alpha$ decay energy, and $h_{\log}$ denotes the hindrance factor associated with unpaired nucleons,
\begin{equation}
h_{\log}=
\begin{cases}
h_{ee}, & Z \ \mathrm{and}\ N \ \mathrm{even},\\
h_{eo}, & Z \ \mathrm{even\ and}\ N \ \mathrm{odd},\\
h_{oe}, & Z \ \mathrm{odd\ and}\ N \ \mathrm{even},\\
h_{oo}, & Z \ \mathrm{and}\ N \ \mathrm{odd}.
\end{cases}
\end{equation}
The ``New2'' parameter set in Ref.~\cite{Peng2009NPR26} is adopted for Eq.~(\ref{eq:A1}), with
$a=1.73539$, $b=-21.3870$, $c=-0.254226$, and $d=-26.5964$.
The corresponding hindrance factors are
$h_{ee}=0$,
$h_{eo}=0.306038$,
$h_{oe}=0.753138$, and
$h_{oo}=0.841341$.	
These parameters were obtained by fitting the experimental $\alpha$ decay energies and half-lives of 71 superheavy nuclei with proton numbers $Z \geq 104$ using the least-squares method. The rms deviation between the logarithms of the calculated and experimental $\alpha$ decay half-lives is 0.5901. 	

The NRDX, UDL, and Royer formulas are given by
	\begin{equation}
	\log_{10}T_{1/2}^{\rm NRDX}
	=a\sqrt{\mu}\,Z_{\alpha}Z_dQ_{\alpha}^{-1/2}
	+b\sqrt{\mu}(Z_{\alpha}Z_d)^{1/2}+c,
	\end{equation}
	\begin{equation}
	\begin{aligned}
	\log_{10}T_{1/2}^{\rm UDL}
	=&\,a\sqrt{\mu}\,Z_{\alpha}Z_dQ_{\alpha}^{-1/2}\\
	&+b\left[\sqrt{\mu}\,Z_{\alpha}Z_d
	(A_{\alpha}^{1/3}+A_d^{1/3})\right]^{1/2}+c,
	\end{aligned}
	\end{equation}
	\begin{equation}
	\log_{10}T_{1/2}^{\rm Royer}
	=(a+bA^{1/6}\sqrt{Z})
	+cZQ_{\alpha}^{-1/2}.
	\end{equation}
}
In these three formulas,
$\mu=A_{\alpha}A_d/(A_{\alpha}+A_d)$, 
$A_{\alpha}$ ($Z_{\alpha}$), and $A_d$ ($Z_d$) denote the mass (proton) numbers of the $\alpha$ particle and daughter nucleus, respectively.
$Q_{\alpha}$ (in MeV) is the $\alpha$ decay energy.
The corresponding parameter sets are taken from Ref.~\cite{Ismail2025CPC49} and summarized in Table~\ref{tab:alpha_para}.

The coefficients of these three formulas were determined separately for even-even, even-odd, odd-even, and odd-odd nuclei by fitting to the experimental half-lives of 573 nuclei with proton numbers in the range $52 \leq Z \leq 118$.
The rms deviations between the calculated and experimental $\log_{10} T_{1/2}$ values are 0.635 for NRDX, 0.629 for UDL, and 0.629 for the Royer formula.
Since the parameters were fitted over a relatively broad range of nuclei, the reliability of these formulas in the superheavy region was further investigated. To this end, the experimental $\alpha$ decay energies and half-lives listed in Ref.~\cite{Peng2009NPR26} were adopted to recalculate the $\alpha$ decay half-lives of 71 superheavy nuclei with $Z \geq 104$ using the NRDX, UDL, and Royer formulas. The resulting deviations are 0.741, 0.674, and 0.756, respectively.

\begin{table}[tb]
	\caption{Parameters of the adopted $\alpha$ decay formulas, taken from Ref.~\cite{Ismail2025CPC49}.}
	\label{tab:alpha_para}
	\footnotesize
	\renewcommand{\arraystretch}{0.95}
	\setlength{\tabcolsep}{4pt}
	
	\begin{ruledtabular}
		\begin{tabular}{llccc}
			Formula & Nuclei & $a$ & $b$ & $c$ \\
			\hline
			
			NRDX~\cite{Ni2008PRC78}
			& e-e & 0.4025 & $-1.4841$ & $-12.4258$ \\
			& e-o & 0.4087 & $-1.3802$ & $-15.3409$ \\
			& o-e & 0.4093 & $-1.4632$ & $-13.4067$ \\
			& o-o & 0.4199 & $-1.4132$ & $-15.5589$ \\
			& All nuclei & 0.4033 & $-1.4274$ & $-13.5814$ \\
			\hline
			
			UDL~\cite{Qi2009PRL103}
			& e-e & 0.4089 & $-0.5949$ & $-21.7141$ \\
			& e-o & 0.4149 & $-0.5545$ & $-23.9417$ \\
			& o-e & 0.4167 & $-0.5905$ & $-22.4893$ \\
			& o-o & 0.4272 & $-0.5709$ & $-24.3210$ \\
			& All nuclei & 0.4098 & $-0.5737$ & $-22.4698$ \\
			\hline
			
			Royer~\cite{Royer2000JPGNPP26}
			& e-e & $-25.7207$ & $-1.1385$ & 1.5823 \\
			& e-o & $-27.9119$ & $-1.0524$ & 1.6066 \\
			& o-e & $-26.6051$ & $-1.1229$ & 1.6118 \\
			& o-o & $-28.5625$ & $-1.0722$ & 1.6514 \\
			& All nuclei & $-26.4730$ & $-1.0913$ & 1.5854 \\
		\end{tabular}
	\end{ruledtabular}
\end{table}

\subsection{Spontaneous fission}

The spontaneous fission half-lives have been estimated using two formulas proposed in Refs.~\cite{KARPOV2012IJMPE21} and \cite{Santhosh2010NPA832}. The parameters of the second formula are taken from Ref.~\cite{Pahlavani2023MPLA38}.

The SF(1)  formula is given by
\begin{equation}
\label{eq:SF1}
\begin{split}
\log_{10}\!\left(T_{1/2}^{\rm SF(1)}/\mathrm{s}\right)
=&\ 1146.44 - 75.3153 \frac{Z^2}{A}
+ 1.63792 \left(\frac{Z^2}{A}\right)^2\\
&
- 0.0119827 \left(\frac{Z^2}{A}\right)^3 \\
&+ B_f \left(7.23613 - 0.0947022 \frac{Z^2}{A}\right) \\
&+
\begin{cases}
0, & Z \text{ and } N \text{ are even},\\
1.53897, & A \text{ is odd},\\
0.80822, & Z \text{ and } N \text{ are odd},
\end{cases}
\end{split}
\end{equation}
where $Z$, $N$, and $A$ denote the proton number, neutron number, and mass number of the parent nucleus, respectively.
$B_f$ (in MeV) is the fission barrier.
Considering the scarcity of experimental data in the superheavy region, not only the experimental data but also the realistic theoretical predictions for the region $100 \leq Z \leq 120$ and $140 \leq N \leq 190$ were included in the fitting procedure to determine the coefficients.
For most nuclei, the values of $T_{\mathrm{SF}}^{\mathrm{exp}}/T_{\mathrm{SF}}^{\mathrm{th}}$ lie within six orders of magnitude, with an average deviation of approximately three orders of magnitude.

The SF(2) formula reads
\begin{equation}
\begin{aligned}
\log_{10}\!\left(T_{1/2}^{\rm SF(2)}/\mathrm{yr}\right)
=&\,
a\frac{Z^2}{A}
+b\left(\frac{Z^2}{A}\right)^2 
+c\left(\frac{N-Z}{N+Z}\right)\\
&+d\left(\frac{N-Z}{N+Z}\right)^2+e,
\end{aligned}
\end{equation}
where $Z$, $N$, and $A$ denote the proton number, neutron number, and mass number of the parent nucleus, respectively. The corresponding
parameters $a$, $b$, $c$, $d$, and $e$ are $a=-5.55719$, $b=0.0573899$,
$c=701.006$, $d=-1643.85$, and $e=50.4771$.
The adjustable parameters of this formula were obtained by least-square fitting with the experimental half-lives for nuclei with $88 \leq Z \leq 104$, yielding a rms deviation of 1.070 in $\log_{10}T_{1/2}$.

\subsection{$\beta$ decay}
\begin{table*}[!t]
	\caption{Parameters of the adopted $\beta$ decay formulas.}
	\label{tab:beta_para}
	\footnotesize
	\begin{ruledtabular}
		
		\begin{tabular}{lccccccccc}
			Formula &
			$a_1$ & $a_2$ & $a_3$ & $a_4$ & $a_5$ &
			$a_6$ & $a_7$ & $a_8$ & $a_9$ \\
			\hline
			$\beta^+$ ~\cite{Sobhani2023CJP85} &
			--0.168569888 &  0.025941442 &  1.746899551 & 38.505487933 &  3.820525023 &
			- & - & - & - \\
			
			$\beta^-(1)$ ~\cite{Sobhani2023CJP85} &
			0.445324552 & --0.163670796 & 6.036983382 & --1.780425712 & --3.984503312 &
			- & - & - & - \\	
			
			$\beta^-(2)$ ~\cite{Zhang2007JoPGNaPP34} &
			$3.37{\times}10^{-4}$ & --0.2558 & 0.4028 & --1.0100 & 0.9039 &
			7.7139 & - & - & -\\
			
			$\beta^-(3)$ ~\cite{Zhou2017SCMA60}&
			3.016 & 3.879 & 1.322 & 6.030 & 1.669 &
			11.09 & 1.07 & --0.935 & --5.398 \\
			
			$\beta^-(4)$ ~\cite{JinGe2024APS73}&
			13.660 & 6.148 & 1.054 & --0.563 & 1.372 &
			2.864 & 6.115 & 3.257 & 1.693 \\
			
			$\beta^-(5)$ ~\cite{Tian2025CPC49}&
			14.608 & 6.164 & 0.545 & 3.985 & 5.882 &
			3.610 & 1.608 & 0.498 & - \\
		\end{tabular}
		
	\end{ruledtabular}
\end{table*}
The semiempirical formulas adopted for estimating $\beta$ decay half-lives (in seconds) are taken from Refs.~\cite{Sobhani2023CJP85,Zhang2007JoPGNaPP34,Zhou2017SCMA60,JinGe2024APS73,Tian2025CPC49}. The corresponding expressions are given below:
\begin{equation}
\log T_{1/2}^{\beta^+,\beta^-(1)}
=
a_1 Z
+a_2 A
+a_3 Q_\beta^{-1/4}
+a_4 I
+a_5 ;
\end{equation} 
\begin{equation}
\log_{10}T_{1/2}^{\beta^-(2)}
=(a_1Z+a_2)N+a_3Z+a_4+\mathrm{Shell}(Z,N),
\end{equation}
where 
\begin{equation}
\begin{aligned}
\mathrm{Shell}(Z,N)
=&\,a_5\Big[
e^{-(N-29)^2/15}+e^{-(N-50)^2/37}\\&
+e^{-(N-85)^2/9} 
+e^{-(N-131)^2/3}
\Big] \\
&+a_6 e^{-[(Z-51.5)^2+(N-80.5)^2]/1.9};
\end{aligned}
\end{equation}
\begin{equation}
\begin{aligned}
\ln T_{1/2}^{\beta^-(3)}
=&\,a_6
+\left(
\alpha^2 Z^2
-5
-a_7\frac{N-Z}{A}
\right)
\ln\!\left(Q_\beta-a_8\delta\right) \\
&+a_9\alpha^2 Z^2
+\frac{1}{3}\alpha^2 Z^2\ln A
-\alpha Z\pi
+\mathrm{Shell}(Z,N),
\end{aligned}
\end{equation}
where
\begin{equation}
\begin{aligned}
\mathrm{Shell}(Z,N)
=&\,a_1 e^{-((N-28)^2+(Z-20)^2)/12}\\&
+a_2 e^{-((N-50)^2+(Z-38)^2)/43} \\
&+a_3 e^{-((N-82)^2+(Z-50)^2)/13}\\&
+a_4 e^{-((N-82)^2+(Z-58)^2)/24} \\
&+a_5 e^{-((N-110)^2+(Z-70)^2)/244};
\end{aligned}
\end{equation}
\begin{equation}
\begin{aligned}
\ln T_{1/2}^{\beta^-(4)}
=&\,a_1
+\left(\alpha^2 Z^2-a_2-a_3 I\right)
\ln\!\left(Q_\beta+m_e c^2-a_4\delta\right) \\
&+\ln\!\left(\frac{2r_0}{\hbar c}\right)\alpha^2 Z^2
+\frac{1}{3}\alpha^2 Z^2 \ln A
-a_5\alpha Z\\
&+\mathrm{Shell}(Z,N),
\end{aligned}
\end{equation}
where \begin{equation}
\begin{aligned}
\mathrm{Shell}(Z,N)
=&\,a_6
e^{-[(N-28)^2+(Z-20)^2]/22}\\&
+a_7
e^{-[(N-50)^2+(Z-40)^2]/33} \\
&+a_8
e^{-[(N-82)^2+(Z-56)^2]/33}\\&
+a_9
e^{-[(N-132)^2+(Z-82)^2]/12};
\end{aligned}
\end{equation}
\begin{equation}
\begin{aligned}
\ln T_{1/2}^{\beta^-(5)}
=&\,a_1
-a_2\ln\!\left(Q_{\beta}+m_e c^2+a_3\delta\right)
+\mathrm{Shell}(Z,N) \\
&-a_8\ln\!\left(
Ze^{-N/Z}+Ne^{-Z/N}+N-Z
\right),
\end{aligned}
\end{equation}
where 
\begin{equation}
\begin{aligned}
\mathrm{Shell}(Z,N)
=&\,a_4 e^{-[(Z-20)^2+(N-24)^2]/30}\\
&+a_5 e^{-[(Z-40)^2+(N-50)^2]/40} \\
&+a_6 e^{-[(Z-56)^2+(N-82)^2]/34}\\&
+a_7 e^{-[(Z-82)^2+(N-132)^2]/11}.
\end{aligned}
\end{equation}
In these formulas, $Z$, $N$, and $A$ denote the proton number, neutron number, and mass number of the parent nucleus, respectively. $Q_\beta$ (in MeV) is the $\beta$ decay energy, $I=(N-Z)/A$ is the isospin asymmetry parameter,  $\delta=(-1)^Z+(-1)^N$ is the odd-even effect term, $m_e$ is the electron mass, and $\alpha$ = 1/137 is the fine structure constant, $r_0 = 1.2\,\mathrm{fm}$. $\mathrm{Shell}(Z,N)$ represent shell-correction term. The corresponding parameters for all adopted formulas are summarized in Table~\ref{tab:beta_para}.

For $\beta^+$ and $\beta^-$(1), 
the mean values of the absolute errors in $\log T_{1/2}$ are 0.348 and 0.382, respectively, based on fits to 872 and 828 nuclei. The corresponding ranges of proton numbers are $7 \leq Z \leq 107$ and $2 \leq Z \leq 101$, respectively.
For $\beta^-$(2), a set of parameters was obtained through a least-squares fit to the experimental data for 351 nuclei far from the $\beta$-stability line, with proton numbers ranging from $Z=5$ to 90. The rms deviation between the calculated and experimental $\log_{10}T_{1/2}$ values is 0.27.
For $\beta^-$(3), the parameters were determined by a least-squares fit to the experimental data for 350 nuclei far from the $\beta$-stability line. The average ratio between the calculated $\beta^-$ decay half-lives and those from experiments is 1.69.
For $\beta^-$(4) and $\beta^-$(5), 
the rms deviations of the logarithms of the calculated half-lives from the experimental data are 0.220 and 0.307, respectively, for nuclei with proton numbers ranging from $Z=8$ to 98.

	\bibliographystyle{apsrev4-1}
	\bibliography{Z124}
	
		\end{document}